\documentclass[12pt,bibliography]{article}
\usepackage{hyperref}
\usepackage{epsfig}
\usepackage{amsfonts}
\usepackage{amsmath}
\usepackage{eufrak}
\usepackage{booktabs}
\usepackage{amssymb}
\usepackage{color} 
\usepackage{pifont}
\usepackage{euscript}
\usepackage{amsbsy}
\usepackage{amsthm}
\usepackage{amscd}
\usepackage{amstext}
\usepackage{verbatim}
\usepackage{cancel}
\usepackage{capt-of}
\usepackage{empheq}
\usepackage{graphicx}
\usepackage{caption}
\usepackage{subcaption}

\newcommand{\be}{\begin{equation}}
\newcommand{\ee}{\end{equation}}
\newcommand{\ba}{\begin{eqnarray}}
\newcommand{\ea}{\end{eqnarray}}
\newcommand{\bi}{\begin{itemize}}
\newcommand{\ei}{\end{itemize}}

\newcommand{\Tr}{{\rm Tr}}

\newcommand{\aslash}[1]{\,\,{\raise.15ex\hbox{/}\mkern-12mu #1}}
\newcommand{\bslash}[1]{\,\,{\raise.15ex\hbox{/}\mkern-9mu #1}}

\renewcommand{\bar}{\overline}
\renewcommand{\tilde}{\widetilde}
\renewcommand{\hat}{\widehat}

\newcommand\lrpar{\raise .8ex\hbox{$^\leftrightarrow$} \hspace{-9pt}
\partial}
\newcommand\lpar{\raise .8ex\hbox{$^\leftarrow$} \hspace{-9pt}
\partial}
\newcommand\rpar{\raise .8ex\hbox{$^\rightarrow$} \hspace{-9pt}
\partial}
\newcommand\lrd{\raise .8ex\hbox{$^\leftrightarrow$} \hspace{-9pt}
\nabla}

\newcommand{\gsim}{\lower.7ex\hbox{$\;\stackrel{\textstyle>}{\sim}\;$}}
\newcommand{\lsim}{\lower.7ex\hbox{$\;\stackrel{\textstyle<}{\sim}\;$}}
\begin{document}

\newcommand{\todo}{{ \color{red}{\bf To Do:}} }

\baselineskip=18pt

\setcounter{footnote}{0}
\setcounter{figure}{0}
\setcounter{table}{0}

\begin{titlepage}

\begin{center}
\vspace{1cm}

{\Large \bf  
Thermodynamics of the BMN matrix model \\ at strong coupling
}

\vspace{0.8cm}

{\large  Miguel S. Costa$ ^\dagger$, Lauren Greenspan$ ^\dagger$, Jo\~ao Penedones$ ^\dagger$, Jorge Santos$ ^{\ddagger,\Diamond}$}

\vspace{.5cm}

{\it  $ ^\dagger$Centro de F\'\i sica do Porto,
Departamento de F\'\i sica e Astronomia\\
Faculdade de Ci\^encias da Universidade do Porto\\
Rua do Campo Alegre 687,
4169--007 Porto, Portugal}
\\
\vspace{.3cm}
{\it  $ ^\ddagger$Department of Physics, Stanford University \\ Stanford, CA 94305-4060, USA}
\\
\vspace{.3cm}
{\it  $ ^\Diamond$Department of Applied Mathematics and Theoretical Physics\\
University of Cambridge, Wilberforce Road\\
Cambridge CB3 0WA, UK}

\end{center}
\vspace{1cm}

\begin{abstract}
We construct the black hole geometry dual to the deconfined phase of the BMN matrix model at strong 't Hooft coupling. We approach this solution from the limit of large temperature where it is  approximately that of the non-extremal D0-brane geometry with a spherical $S^8$ horizon. This geometry preserves the $SO(9)$ symmetry of the matrix model trivial vacuum. As the temperature decreases the horizon becomes deformed and breaks the $SO(9)$ to the $SO(6)\times SO(3)$ symmetry of the matrix model. When the black hole  free energy crosses zero the system undergoes a phase transition to the confined phase described by a Lin-Maldacena geometry. We determine this critical temperature, whose computation is also within reach of Monte Carlo simulations of the matrix model.
\end{abstract}

\bigskip
\bigskip

\end{titlepage}

\tableofcontents

\section{Introduction}

Some quantum mechanical systems admit a parametric limit in which they are well described by a classical gravitational theory. Such systems are examples of quantum theories of gravity. 
It is not easy to find a system with this property but the gauge/gravity duality offers  several cases \cite{Maldacena, Itzhaki:1998dd}.

Perhaps, the most striking example is $(0+1)$-dimensional $SU(N)$ Super Yang-Mills (SYM) theory.
This theory contains a finite number  of bosonic and fermionic degrees of freedom which are naturally organized in $N$ by $N$ traceless hermitian matrices $X^i$ and $\Psi^\alpha$, respectively.
This model is often termed BFSS \cite{Banks:1996vh}, with  action  given by
\be
S_{D0}=\frac{N}{2\lambda} \int dt \,\Tr \left[
  (D_t X^i)^2 + \Psi^\alpha D_t \Psi^\alpha
+\frac{1}{2} \left[X^i,X^j\right]^2 + i \Psi^\alpha \gamma_{\alpha \beta}^j [\Psi^\beta,X^j]
\right],
\label{D0action}
\ee 
where $D_t=\partial_t-i[A,\ ]$ is the covariant derivative and summation over spatial indices 
$i,j=1,\dots,9$ and  spinor indices $\alpha,\beta=1,\dots,16$ is implicit. 
By dimensional analysis, one concludes that the 't Hooft coupling $\lambda$ has units of energy cubed. Therefore, the thermodynamics of this system is controlled by two dimensionless parameters: $N$ and $\tau=T/\lambda^{\frac{1}{3}}$, where $T$ is the temperature.
According to the gauge/gravity duality, at large $N$ and small dimensionless temperature 
$\tau$ this theory is dual to 11-dimensional supergravity in the 
following black hole geometry\footnote{This follows from the decoupling limit of $N$ coincident D0-branes in type IIA supergravity \cite{Itzhaki:1998dd}, which
leads to a charged spherically symmetric black hole in ten dimensions. 
From our point of view, it is more convenient to work with the  solution uplifted to eleven dimensions, where it is purely geometric and
describes a black string with horizon topology $S^1\times S^8$.
In our conventions, the 11-dimensional Newton constant is given by $16\pi G_N= (2\pi)^8 g_s^3 \ell_s^9$ and the periodic coordinate $z$ obeys $z\sim z+2\pi g_s \ell_s$.
}
\be
ds^2= \frac{dr^2}{f(r)}+r^2 d\Omega_8^2+
\frac{R^7}{r^7}dz^2+f(r)
dt \left(2dz-\frac{r_0^7}{R^7}dt 
\right),
\label{D011D}
\ee
where 
\be
f(r)=1-\left(\frac{r_0}{r}\right)^7\,,\ \ \ \ \ \ \ \ 
\left(\frac{R}{\ell_s}\right)^7=60\pi^3g_s N\,,\ \ \ \ \ \ \ \ \ 
\left(\frac{r_0}{\ell_s}\right)^5=\frac{120\pi^2}{49} \left(2\pi g_s N\right)^{\frac{5}{3}}\tau^2\,,
\label{ParameterRelations}
\ee
with $\ell_s$ the string length and 
$g_s=4\pi^2  \ell^3_s\lambda/N$ the string coupling.

This gravitational description
leads to the following prediction for the large $N$ and low temperature expansion of the free energy\footnote{We are assuming 
$\tau\gg N^{-\frac{5}{9}} $.
 For lower temperatures, the black string suffers from the Gregory-Laflamme instability and the stable black hole should have $S^9$ horizon topology. 
For even lower temperatures $\tau \sim N^{-\frac{5}{6}}$, the curvature at the horizon of (\ref{D011D}) reaches the Planck scale.
}
\be
\frac{F}{\lambda^{\frac{1}{3}}N^2}=
\left[c_1 \tau^\frac{14}{5} + c_2 \tau^\frac{23}{5} +\dots\right]+
\frac{1}{N^2}\left[c_3 \tau^\frac{2}{5} + c_4 \tau^\frac{11}{5} +\dots\right]+
\dots\,,
\label{FoftauandN}
\ee
where the $c_i$ are numerical coefficients.
The leading term follows from the classical black hole thermodynamics of (\ref{D011D}), which gives $c_1=-\frac{1}{21} \big(\frac{120\pi^2}{49} \big)^{7/5}$.
The coefficient $c_2$ is not known analytically because it follows from unknown $\alpha'^{\,3}$ corrections to type IIA effective action \cite{Hanada:2008ez}.
The $1/N^2$ terms correspond to quantum corrections associated to string loops in the 10 dimensional picture. 
Notably, the coefficient $c_3$ has been computed recently using the quartic curvature corrections to 11D supergravity \cite{Hyakutake2013}.
It is an outstanding challenge to reproduce analitically these predictions directly from the matrix quantum mechanics (\ref{D0action}). In fact, the state of the art is a scaling hypothesis for the several terms of the moduli effective action that correctly predicts the leading low-temperature dependence $\tau^{\frac{14}{5}}$ but it is unable to fix the coefficient $c_1$ and any of the subleading terms
\cite{Wiseman2013, Morita:2013wfa}.
The mean field approximation of \cite{Kabat:1999hp, Kabat:2000zv, Kabat:2001ve}
claimed partial success in reproducing the gravity prediction but their numerical method breaks down for sufficiently low temperature \cite{Lin:2013jra}.

In a remarkable series of papers \cite{Catterall:2007fp,  Anagnostopoulos:2007fw, Catterall:2008yz, Hanada:2008ez,  Catterall:2009xn, Hanada:2011fq, Hanada:2013rga}, 
the authors performed Monte-Carlo simulations of the matrix quantum mechanics (\ref{D0action}) at finite temperature.
In \cite{Hanada:2008ez}, they studied the planar limit ($N\to \infty$) at low temperature and obtained the first two terms in equation (\ref{FoftauandN}). Their results agree with all available analytical results from the gravity dual and provide a prediction for $c_2$. 
More recently \cite{Hanada:2013rga},  a study of 
$1/N^2$ effects confirmed the gravitational prediction for the coefficient $c_3$.
This is among the most impressive tests of the gauge/gravity duality we are aware of. Notice that this includes 
quantum gravity loop effects and  probes the regime of chaotic dynamics where supersymmetry is completely broken and integrability is absent. 

In this paper, we study the thermodynamics of a massive deformation of the matrix quantum mechanics (\ref{D0action}). This model 
goes by the name of PWMM (plane wave matrix model)
or BMN after the authors of \cite{BMN}.
Its action reads
\be
S=S_{D0}-\frac{N}{2\lambda} \int dt \,\Tr \left[
\frac{\mu^2}{ 3^2} ( X^i)^2 +
\frac{\mu^2}{ 6^2} (X^a)^2 +
 \frac{\mu}{4}\Psi^\alpha \left(\gamma^{123}\right)_{\alpha \beta} \Psi^\beta 
+i\frac{2\mu}{3} \epsilon_{ijk} X^iX^jX^k \right] ,
\label{PWMMaction}
\ee 
where the indices $i,j,k$ run over $1,2,3$ and the index $a$ runs over $4,\dots,9$.
The matrix $\gamma^{123}$ is equal to $ \frac{1}{6} \epsilon_{ijk} \gamma^i\gamma^j \gamma^k$, with $\epsilon_{ijk}$ the standard 3 dimensional $\epsilon$-tensor.
This means that the mass parameter $\mu$ breaks the $SO(9)$ global symmetry of (\ref{D0action}) down to $SO(6)\times SO(3)$. The deformation also retains maximal supersymmetry \cite{BMN}.

The BMN model has three significant advantages over the BFSS model. The first is that it has a discrete energy spectrum and a well defined canonical ensemble.
Notice that, strictly speaking, the canonical ensemble of the matrix quantum mechanics (\ref{D0action})  does not exist.
\footnote{In \cite{Catterall:2009xn}, it was shown explicitly that the free energy of BFSS has an infrared divergent contribution at order $N$.
The same paper suggested the study of the BMN model as a way to tame this problem.
}
 The reason for this is that the eigenvalues of commuting matrices $X^i$ can be made arbitrarily large without   energy cost.
In fact, the Monte-Carlo simulations  work  because there is a meta-stable thermal equilibrium with a decay rate that is very small at large $N$.
The second advantage is that the BMN model has a dimensionless coupling constant $g\equiv \lambda/\mu^3$ that,  together with the dimensionless temperature $T/\mu$, parametrize  a two-dimensional phase diagram.
This means that we can use the dual gravitational description at large $N$ and strong coupling $g\gg 1$, to predict many observables as functions of the dimensionless temperature  $T/\mu$.
Finally, the third advantage is  that the BMN model is expected to have a phase transition whose critical temperature  should be easy to measure in Monte-Carlo simulations.
\footnote{This is qualitatively similar to the case of 2D SYM compactified on a circle \cite{Catterall:2010fx}.}

\begin{figure}[h]
\centering
\includegraphics[height = 0.3\textheight]{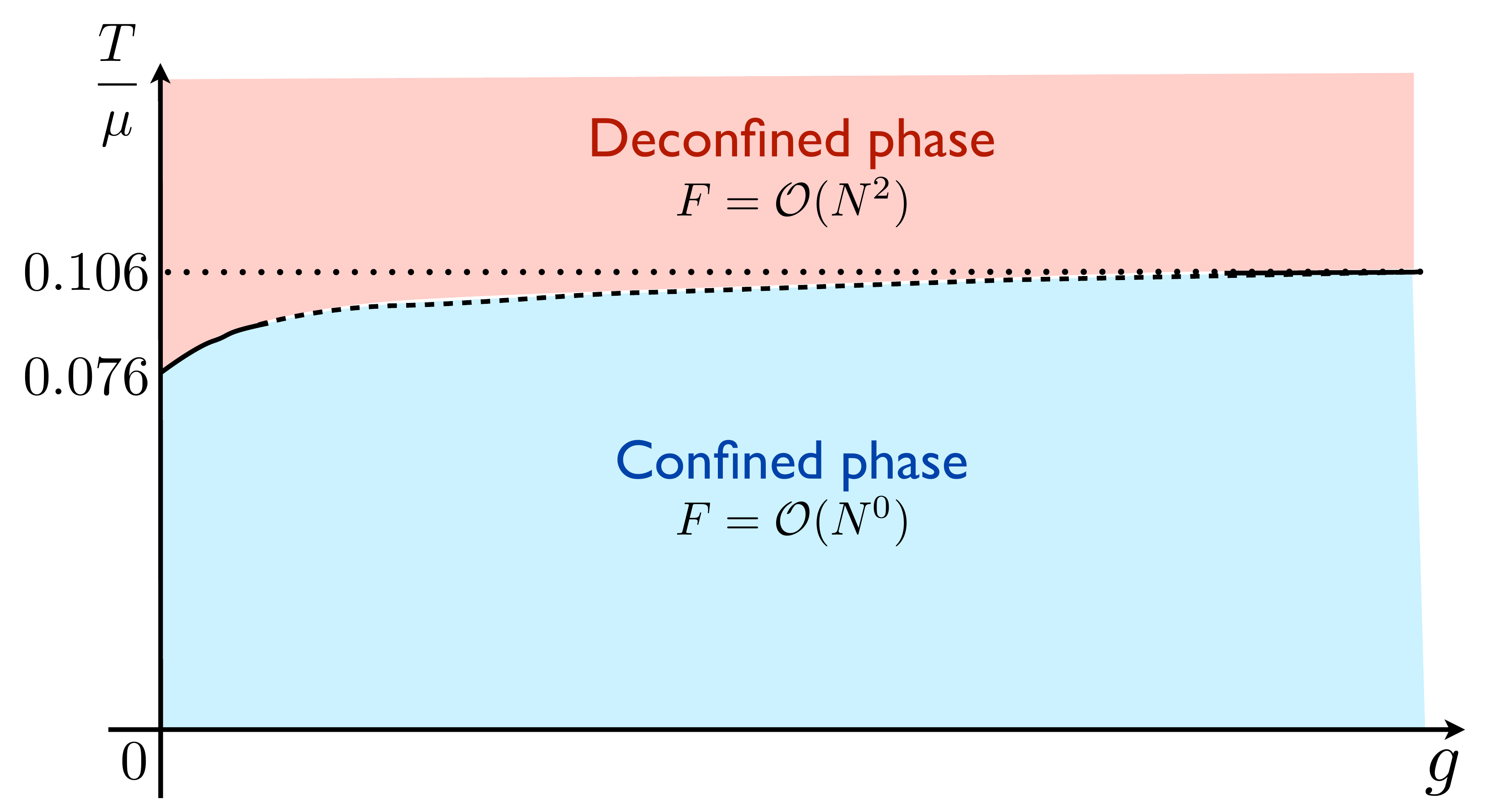}
\caption{\label{phasediagram}
The  phase diagram of the PWMM. At high temperature, the system is in a deconfined phase where the free energy scales like $N^2$. As we lower the temperature, the system undergoes a first order phase transition to a confined phase where the free energy scales as $N^0$.
The critical temperature $T_c$ depends on the dimensionless coupling $g$ and it can be computed 
in perturbation theory for $g \ll 1$.
In this paper we determine 
$T_c$  at strong coupling from the study of the black hole dual to the deconfined phase of the PWMM.}
\end{figure}

In figure \ref{phasediagram} we depict the phase diagram of the theory  in the planar limit $N\to \infty$.
In the weak coupling regime $g \ll 1$, the dynamics of the system can be studied using perturbation theory.  
One starts by expanding the fields around one of the minima of the potential
\be
\frac{N}{2\lambda} \, \Tr \left[ 
\frac{\mu^2}{ 6^2} (X^a)^2  
-\frac{1}{2} \left[X^a,X^b\right]^2
-\frac{1}{2}\left( [X^i,X^j]-i\frac{\mu}{3} \epsilon_{ijk}  X^k \right)^2
\right].
\ee
Since this is a sum of squares, the minima are given by $X^a=0$ and $X^i=\frac{1}{3}\mu\, J^i$, with $J^i$  a $N$-dimensional representation  of $SU(2)$ (in other words $[J^i,J^j]=i\epsilon_{ijk} J^k$).
This means that the minima are in one-to-one correspondence with integer partitions of $N$ because we can form an $N\times N$ block diagonal matrix by adding many blocks with $SU(2)$ irreducible representations.
In the large $N$ limit, tunnelling between different vacua is suppressed and it is possible to study the thermodynamics associated to each minimum \cite{Kawahara:2006hs}.\footnote{More precisely, this works for vacua that are associated with a reducible representation of $SU(2)$ that contains many copies (of order $N$) of a few irreducible representations of fixed dimension.  
If the dimensions of the irreducible representations scale with $N$ and there are a fixed number of them (membrane  states) then the free energy is of order 1 and there is no phase transition. In fact, the fluctuations around these vacua become free in the 't Hooft limit \cite{Furuuchi:2003sy}.
}
In this paper, we will focus on the trivial vacuum $X^a=X^i=0$. The excitations above this vacuum are gapped and weakly coupled if $g\ll 1$. For energies much greater than $\mu$ and much smaller than $\mu N^2$ the density of states grows exponentially with energy. This leads to a Hagedorn phase transition at $T=\frac{\mu}{12\log 3}$ for $g=0$. At weak coupling $g$, this becomes a first order phase transition whose critical temperature can be computed in perturbation theory
 \cite{Furuuchi:2003sy, Hadizadeh:2004bf}
\be
 T_c(g)  = \frac{\mu}{12\log 3}
\left[
1+\frac{2^6\cdot 5}{3} g
-\left(\frac{23\cdot19927}{2^2\cdot3}+
\frac{1765769}{2^4\cdot3^2}\log 3 \right) g^2
+\mathcal{O} (g^3 )
\right].
\ee
We call the high temperature phase the deconfined phase because  the free energy scales as $N^2$. For $T<T_c$ the system is in the confined phase where the free energy scales as $N^0$.

The main goal of the present work is to determine the fate of this phase transition at strong coupling $g \gg 1$.
It is instructive to compare figure \ref{phasediagram} with the phase diagram  of $\mathcal{N}=4$  SYM on $S^3$. In this comparison $1/\mu$ plays the role of the radius of $S^3$ and $g$ plays the role of the 4-dimensional 't Hooft coupling.\footnote{In fact, the action (\ref{PWMMaction}) can be obtained from the action of SYM on $S^3$ by truncating the 4-dimensional fields to their zero modes (more precisely, projecting to $SU(2)_L$ invariant modes) \cite{Kim:2003rza}.
} 
The 4-dimensional theory also has a first order phase transition that starts with a Hagedorn transition of the free theory \cite{Aharony:2003sx}. At strong coupling, this transition corresponds to the Hawking-Page transition in the dual $AdS_5$ gravitational description \cite{Hawking:1982dh, Witten:1998zw}. We will argue that the   PWMM has a very similar phase diagram. In particular, we will find a Hawking-Page like phase transition in the dual gravitational description of the PWMM and predict the strong coupling limit of the critical temperature,
\be
\lim_{g\to \infty} \frac{T_c(g)}{\mu} = 0.105905(57) \,.
\ee
It would be remarkable to  confirm this prediction with Monte-Carlo simulations of the PWMM at strong coupling.
We believe this  to be accessible with the methods of \cite{ Catterall:2007fp,  Anagnostopoulos:2007fw, Catterall:2008yz, Hanada:2008ez, Hanada:2013rga}.

The dual geometries  to each vacuum of the PWMM were constructed in \cite{Lin:2004nb, Lin:2005nh}. These SUSY vacuum geometries, including the one dual to the trivial vacuum, are surprisingly complicated (see appendix \ref{ap:LinMalda})\cite{Ling:2006up}.
Nevertheless, they share an important feature in that  they asymptote to the plane wave solution of M-theory
\be
ds^2= dx^i dx^i+dx^a dx^a+
2dt dz-
\left(\frac{\mu^2}{3^2} x^i x^i +  
\frac{\mu^2}{6^2} x^a x^a
\right) dt^2\,,\ \ \ \ \ \ 
dC=\mu\,dt\wedge dx^1 \wedge dx^2 \wedge dx^3\,.
\label{PW11D}
\ee
Fortunately, we will not need the detailed form of these vacuum geometries. Our strategy will be to start from very high temperature ($T\gg \mu$) and gradually decrease it.  This means that our starting solution is the uplifted 11D SUGRA solution (\ref{D011D}) for which the 4-form field strength vanishes.
This geometry has the same $SO(9)$ symmetry of the trivial vacuum $X^a=X^i=0$. 
We will then continuously deform this solution by turning on a non-normalizable mode of $dC$ that corresponds to the relevant deformation that takes the BFSS to the BMN model. This deformation breaks the $SO(9)$ symmetry of (\ref{D011D}) to $SO(6)\times SO(3)$,  making the field equations analytically intractable. 
 In the next section, we explain how this is done in detail, including the numerical methods to solve the relevant Einstein equations.
In section \ref{sec:thermo}, we determine the free energy of the black hole constructed in section \ref{sec:BH} and the strong coupling limit of the critical temperature $T_c(g)$.
We also calculate thermal expectation values of several operators in the high temperature deconfined phase.
We conclude in section \ref{sec:disc} with a discussion and comments about open questions.

\section{Deformed Black Hole  \label{sec:BH} }

Let us start by fixing our conventions for the bosonic piece of the 11-dimensional SUGRA action
\begin{align}
I&=\frac{1}{16 \pi G_N}   \int \left(  \eta \mathcal{R} + \frac{1}{2} dC\wedge \star dC -\frac{1}{6}   C \wedge dC \wedge dC\right),
\label{11DSUGRA}
\end{align}
where  $\eta$ is the space-time volume form, $\mathcal{R}$ is the Ricci scalar and 
$C$ is a 3-form gauge potential. 
Any stationary solution compatible with the $SO(6)\times SO(3)$ global symmetry and invariant under translations along the eleventh direction can be written as
\begin{align}
ds^2=& -A\,\frac{(1-y^7)  }{y^7}d\eta^2
+T_4\,
y^7  \left[ d\zeta+\Omega\,\frac{(1-y^7)    d\eta}{y^7}\right]^2
\nonumber \\
&+
\frac{1}{y^2}\left[B\,
\frac{\left(dy+F dx \right)^2}{(1-y^7)y^2 }+
T_1\,\frac{4dx^2}{2-x^2}+T_2\,x^2(2-x^2) d\Omega_2^2+T_3\,
(1-x^2)^2 d\Omega_5^2
\right] ,\label{11Dansatz}
\\
C=& \left(M \, d\eta +L\,d\zeta\right) \wedge d\Omega_2 \,,
\nonumber
\end{align}
where  $\zeta \sim \zeta +2\pi$ is the periodicity of the 11-dimensional circle and  the functions  
 $A$, $B$, $F$, $\Omega$, $T_1$, $T_2$, $T_3$, $T_4$, $M,L$ 
depend on the radial coordinate $0 \le y < 1$ and on the angular coordinate $0 \le x \le 1$.
We shall see that $y=1$ corresponds to the black hole horizon and  $y=0$ to the 
asymptotic region, which matches the plane wave geometry (\ref{PW11D}).
The  angular coordinate $x$  was introduced to break the $SO(9)$ symmetry of an 
eight sphere to  $SO(6)\times SO(3)$. We can think of
$x=0$ as the $S^5$ {\em equator} and $x=1$ as the $S^2$ {\em pole},
of the 8-dimensional surface $d\eta=d\zeta=dy=0$. 
This form of the solution is tailored to the numerical methods we will use.
In particular, all quantities are dimensionless and the domain of the unknown functions is the unit square. The physical solution
can then be obtained by using the scalings of the 11D SUGRA action under the following transformations
\begin{align}
g_{ab}\rightarrow \lambda^2 g_{ab}\,, \ \ 
C_{abc}\rightarrow \lambda^3 C_{abc} \ \ \ \ &\Rightarrow\ \ \ 
I\rightarrow \lambda^9 I\,,\nonumber
\\
 \label{eq:trans}
\\
 \zeta \sim \zeta + 2\pi \ \rightarrow \  \zeta \sim \zeta + 2\pi \lambda' \ \ \ \ &\Rightarrow\ \ \ I\rightarrow\lambda' I\,,\nonumber
\end{align} 
More concretely, the  physical solution will be obtained from (\ref{11Dansatz}) 
by multiplying the metric by $r_0^2$ and the 3-form $C$ by $r_0^3$, and by changing the period of the non-contractible M-theory circle according to
 \be
 \zeta \sim \zeta + 2\pi \left(\frac{R}{r_0}\right)^{\frac{7}{2}}\frac{g_s\ell_s}{r_0}\,.
 \label{ScaleX11}
 \ee
Both operations are symmetries of the 
equations of motion, but change the value of the on-shell action to
\be
I= \frac{r_0^9}{16 \pi G_N}
\left(\frac{R}{r_0}\right)^{\frac{7}{2}}\frac{g_s\ell_s}{r_0}\, \hat{I}=
\frac{15}{28} \left(\frac{15}{14^2\pi^8}\right)^{\frac{2}{5}} N^2 \tau^{\frac{9}{5}}\,\hat{I}\,,
\label{physAction}
\ee
where we defined the dimensionless action $\hat{I}$ to be the 11D SUGRA action (\ref{11DSUGRA}) evaluated on the Ansatz (\ref{11Dansatz}) and stripped of the overall factor of $1/(16 \pi G_N)$.

In the last equality of (\ref{physAction}), we used the relations (\ref{ParameterRelations}) between the gravitational parameters and the variables of the dual matrix quantum mechanics.
When computing the action of a solution, care must be taken by adding boundary terms that renormalise the on-shell action. 
In what follows we shall assume that such  counter-terms preserve both of the scaling operations described above.
Similarly, the {\em physical} Bekenstein-Hawking entropy becomes
\begin{equation}
S=\frac{r_0^9}{4G_N}
\left(\frac{R}{r_0}\right)^{\frac{7}{2}}\frac{g_s\ell_s}{r_0} \,\hat{S}=
\frac{15\pi}{7} \left(\frac{15}{14^2\pi^8}\right)^{\frac{2}{5}} N^2 \tau^{\frac{9}{5}}\hat{S}\,,
\label{physEntropy}
\end{equation}
where $\hat{S}$ is the dimensionless horizon area computed with the metric (\ref{11Dansatz}),
explicitly given by
\be
\hat{S}=\int_{\cal H} d^9x \sqrt{h}=
16 \pi^5 \int_0^1 dx\, x^2 (1-x^2)^5
\left[ (2-x^2) T_1(1,x) T_2^2(1,x) T_3^5(1,x) T_4(1,x) \right]^{\frac{1}{2}} \,,
\label{Shat}
\ee
where $h$ is the determinant of the induced metric on the horizon, which has $S^8\times S^1$ topology. 

To see how this works in practice for a simple case, consider the exact solution given by 
 $A=B=\Omega=T_1=T_2=T_3=T_4=1$ and $F=M=L=0$.
Changing coordinates,
\be
y=\frac{r_0}{r}\,,\ \ \ \ \ \ \ 
\eta= \left(\frac{r_0}{R}\right)^{\frac{7}{2}}
\frac{t}{r_0} \,,\ \ \ \ \ \ \ \ 
\zeta= \left(\frac{R}{r_0}\right)^{\frac{7}{2}}
\frac{z}{r_0} \,,
\label{coordchange}
\ee 
and multiplying the metric  (\ref{11Dansatz}) by $r_0^2$ one recovers the 11-dimensional uplift of the non-extremal D0-brane solution (\ref{D011D}). Notice that after the Wick rotation $\eta \to i \eta$, the  Euclidean time circle of  (\ref{11Dansatz}) must have period $4\pi/7$ in order to avoid a conical singularity. Using (\ref{coordchange}) this fixes the
periodicity of the dimensionfull Euclidean time, which
is consistent with the relations 
(\ref{ParameterRelations}) between the temperature and the parameter $r_0$. Moreover, using the dimensionless area of the horizon  
$\hat{S}= 2\pi\, {\rm Vol}(S^8)$, we obtain\footnote{Notice that this is compatible with the first term of  (\ref{FoftauandN}) and the first law of thermodynamics $\frac{\partial F}{\partial T}=-S$.
}
\be
S= 
\frac{2}{15} \left(\frac{120\pi^2}{49} \right)^{\frac{7}{5}} N^2 \tau^{\frac{9}{5}} \,.
\ee

This exact solution describes the high temperature limit $T/\mu \to \infty$ of the PWMM. To lower the temperature, we need to appropriately turn on the 3-form potential $C$.  This is implemented in the Ansatz  (\ref{11Dansatz}) by requiring the
function $M=M(x,y)$ to have the
 following asymptotic behaviour
\be
M \approx \hat{\mu}\, \frac{ x^3(2-x^2)^{\frac{3}{2}}}{y^3}\,,\ \ \ \ \ \ \ y \to 0\,.
\label{Mbc}
\ee
To find out the physical meaning of the parameter $\hat{\mu}$, we compute the asymptotic behaviour of the physical field strength $dC$,
determined after multiplying by $r_0^3$ and changing coordinates as in (\ref{coordchange}),
\be
dC \approx  \frac{12\pi}{7} \,\hat{\mu} \,T \,
dt\wedge d(rx\sqrt{2-x^2})\wedge r^2x^2(2-x^2) d\Omega_2\,.
\label{asymptotics_dC}
\ee
Identifying $rx\sqrt{2-x^2}$ as the radial coordinate on the 3-plane that contains the 2-sphere, and comparing with the M-theory plane wave solution (\ref{PW11D}), we conclude that
\be
\hat{\mu}=\frac{7}{12\pi}  \frac{\mu}{T}\,.
\label{mu_hat}
\ee

In section \ref{sec:bc} below, we will explain the precise boundary conditions that uniquely fix   the solution. However, the intuition is clear: we require regularity at the axes of symmetry $x=0$, $x=1$ and $y=1$. 
In particular, the Euclidean period of the $\eta$ coordinate is always $4\pi/7$ because we impose $A=B$ at the horizon $y=1$.\footnote{Recall that in Euclidean signature the horizon is the fixed point of time translation symmetry.}
At infinity ($y\to0$), we impose that 
$A,B,\Omega,T_1,T_2,T_3,T_4 \to 1$, that $F,L \to 0$,   and (\ref{Mbc}). In this way, we obtain a one parameter family of (dimensionless) solutions parametrized by $\hat{\mu}$. The physical entropy of the system, for example, is then computed using (\ref{physEntropy}).
Notice that this agrees precisely with the free energy scaling predicted in \cite{Wiseman2013} from the assumption that the tree level and 1-loop contributions for the moduli effective action are of the same order in the strongly coupled regime.
It is also clear that thermal expectation values that are non-zero at $\mu=0$ (\emph{i.e.} in the non-extremal D0-brane) get multiplied by a function of $\hat{\mu}$, again in agreement with  \cite{Wiseman2013}.

\subsection{Harmonic Einstein equations - DeTurck method}
Our approach to solving Einstein's equations, the so called DeTurck method, was first introduced in \cite{Headrick:2009pv} and studied in great detail in \cite{Figueras:2011va}. Its generalization to finding stationary solutions of the form discussed in this manuscript was first detailed in \cite{Adam:2011dn}.

We first note that in both the line element and gauge field ans\"atze (\ref{11Dansatz}), we have partially gauge fixed coordinate invariance and completely gauge fixed the gauge redundancy associated with $C\to C+\mathrm{d}\Lambda$, where $\Lambda$ is a two-form\footnote{From the perspective of the gravitational system, the functions $M$ and $L$ behave as scalar fields under arbitrary reparametrizations of $x$ and $y$, meaning that no gauge fixing procedure is necessary for these matter fields.}. However, the line element (\ref{11Dansatz}) still exhibits \emph{full} diffeomorphism invariance for arbitrary reparametrizations of $x$ and $y$, which we will fix using the DeTurck method.

The method can be best understood if we write the 11-dimensional Einstein's equations in the trace reversed form
\begin{equation}
\tilde{E}_{ab} \equiv R_{ab} - \frac{1}{12}\left(G_{acde}G_{b}^{\phantom{b}cde}-\frac{g_{ab}}{12}G^{cdef}G_{cdef}\right)=0\,,
\label{EinsteinEquation}
\end{equation}
where $G=dC$ is the field strength.
The idea is to solve for $E_{ab}\equiv \tilde{E}_{ab}-\nabla_{(a}\xi_{b)}=0$, where $\xi^a = [\Gamma_{bc}^a(g)-\Gamma^{a}_{bc}(\bar{g})]g^{bc}$, $\Gamma(\mathfrak{g})$ is the Levi-Civita connection associated with a metric $\mathfrak{g}$ and $\bar{g}$ is a reference metric. The reference metric $\bar{g}$ is chosen to have the same asymptotic and conformal structure as the metric we want to determine, \emph{i.e.} the metric $g$. Here, our reference metric is just given by the line element (\ref{11Dansatz}) with $A = B = T_1=T_2=T_3=T_4=\Omega=1$ and $F=0$. One can show that the resulting system of equations obtained via $E_{ab}=0$ and $d\star dC=0$ is of the form $\gamma^{ab}\partial_a \partial_b Q_i+F_i(\partial_a Q_k,Q_k)=0$, where $F_i$ is a complicated function of $\{Q_i\} =\{A,B,F,T_1,T_2,T_3,T_4,\Omega,M,L\}$ and their first derivatives along $x$ and $y$, and $\gamma^{ab}$ is a two-dimensional positive symmetric matrix ($\gamma^{ab}$ is the inverse of the metric tensor (\ref{11Dansatz}) restricted to a $x,y$ plane with all the other coordinates fixed). This means that, under the appropriate boundary conditions, which we shall discuss below, $E_{ab}=0$ forms a system of Elliptic partial differential equations.

It is clear that any solution to $\tilde{E}_{ab}=0$ with $\xi=0$ is a solution to $E_{ab}=0$, however, the converse is not necessarily true. Under some special circumstances, and for certain types of matter fields, one can show that solutions with $\xi\neq0$, coined Ricci solitons, cannot exist \cite{Figueras:2011va}. However, the case under consideration is not under this special class. Fortunately, if the system of partial differential equations $E_{ab}=0$ is Elliptic, it can be solved as a boundary value problem for well-posed boundary conditions and the solutions should be locally unique. This means that an Einstein solution cannot be arbitrarily close to a soliton solution and one should easily be able to distinguish the Einstein solutions of interest from solitons by monitoring $\xi$. This is the approach we are going to undertake here. However, we first need to show that our boundary conditions give rise to an Elliptic problem, and that they are consistent with $\xi=0$.

\subsection{Boundary conditions\label{sec:bc}}
Our solution naturally lives on a square grid, with $x=0$ denoting the fixed points of the $SO(3)$ symmetry, $x=1$ the fixed points of the $SO(6)$, $y=1$ denoting the horizon location and $y=0$ the conformal boundary.
In an abuse of language, we shall refer to $x=0$ as the $S^5$ {\em equator}, and $x=1$ as the $S^2$ {\em pole}, of the 8-dimensional surface
$d\eta=d\zeta=dy=0$.

We will be interested in measuring certain quantities near the conformal boundary located at $y=0$. 
The leading order term in the  metric near the boundary is just that of the D0-brane, since the matrix model massive deformation is irrelevant in the UV. Thus we will have
\begin{align}
&A(x,y)=B(x,y)=\Omega(x,y)=T_1(x,y)=T_2(x,y)=T_3(x,y)=T_4(x,y)=1+O(y)\,,
\nonumber\\
&F(x,y)=O(y)\,.
\end{align}
This guarantees that, asymptotically,  the 8-dimensional surface
$d\eta=d\zeta=dy=0$ becomes a round $S^8$ and that the total D0-brane charge is fixed.
The leading term for the functions $M$ and $L$ is fixed by requiring that the non-normalizable mode dual to the massive
deformation of the matrix model is turned on.
At leading order in $y$ it suffices to consider linear perturbations of the 3-form potential $C$, which then fix the leading behaviour in $y$ of the functions $M$ and $L$.
These perturbations are naturally expanded in a basis of harmonics of the asymptotic $S^8$.
For the  particular Ansatz (\ref{11Dansatz}) for  the 3-form potential,
\be
C= \left(M \, d\eta +L\,d\zeta\right) \wedge d\Omega_2 \,,
\ee 
the decomposition in $S^8$ harmonics only  includes the harmonic 2-forms
\be
\omega_l=  \mathbb{H}_l(x) d\Omega_2\,,
\ee
with $l$ odd and
\be
\mathbb{H}_l(x)=x^3\! \left(2 - x^2\right)^{3/2}\,_2F_1\!\left(\frac{1}{2}-\frac{l}{2},4+\frac{l }{2},\frac{5}{2},x^2 \left(2-x^2\right)\right),
\ee
where $\,_2F_1$ is a hypergeometric function.
These harmonic 2-forms $\omega_l$ satisfy
\be
\star_8 d (\star_8 d \omega) = - \lambda_l \omega\,, \quad\quad\lambda_l =   (l+2)(l+5) \,,
\ee
where $\star_8$ is the Hodge dual on $S^8$. Note that for odd $l$, $\omega_l$ is  invariant under the action of the  $SO(6) \times SO(3)$ subgroup of the $SO(9)$ isometry of the 
$S^8$.
Finally, analysing the perturbations of the 3-form potential $C$, one concludes that the required non-normalizable mode 
has non-vanishing functions $M$ and $L$, and 
can be written in terms of the harmonic form $\omega_l$ with $l=1$.
Asymptotically, this  fixes their leading behaviour  to be given by
\be
M(x,y)=\hat{\mu} y^{-3} \mathbb{H}_1(x)
+O(y^{-2})\,,\quad\quad
L(x,y)=\frac{3}{2}\hat{\mu} y^{4} \mathbb{H}_1(x)
+O(y^{5})\,.
\label{Q9Q10_bc}
\ee 

To better understand the boundary conditions at the   conformal boundary we actually need to consider in more detail the asymptotic expansion of the fields. We now turn to this problem.

\subsubsection{Asymptotic expansion   at conformal boundary
\label{sec:asympAnsatz}
}
In general, each function will have an asymptotic expansion in powers of $y$. For example,
\be
A(x,y)= \sum_n A_n(x) y^n\,.
\ee
The equations of motion yield second order coupled differential equations  in the variable $x$ for all the  coefficient functions like  $A_n(x)$. These can be easily solved assuming that only smooth solutions on the $S^8$ of the
boundary are allowed, that is to say, all the coefficient functions   admit an  expansion in harmonics on the $S^8$ of the boundary. These harmonics can be of 
scalar, vector or tensor type  and must be invariant under the unbroken $SO(3)\times SO(6)$ symmetry.

\begin{figure}[t]
\centering
\includegraphics[height = 0.35\textheight]{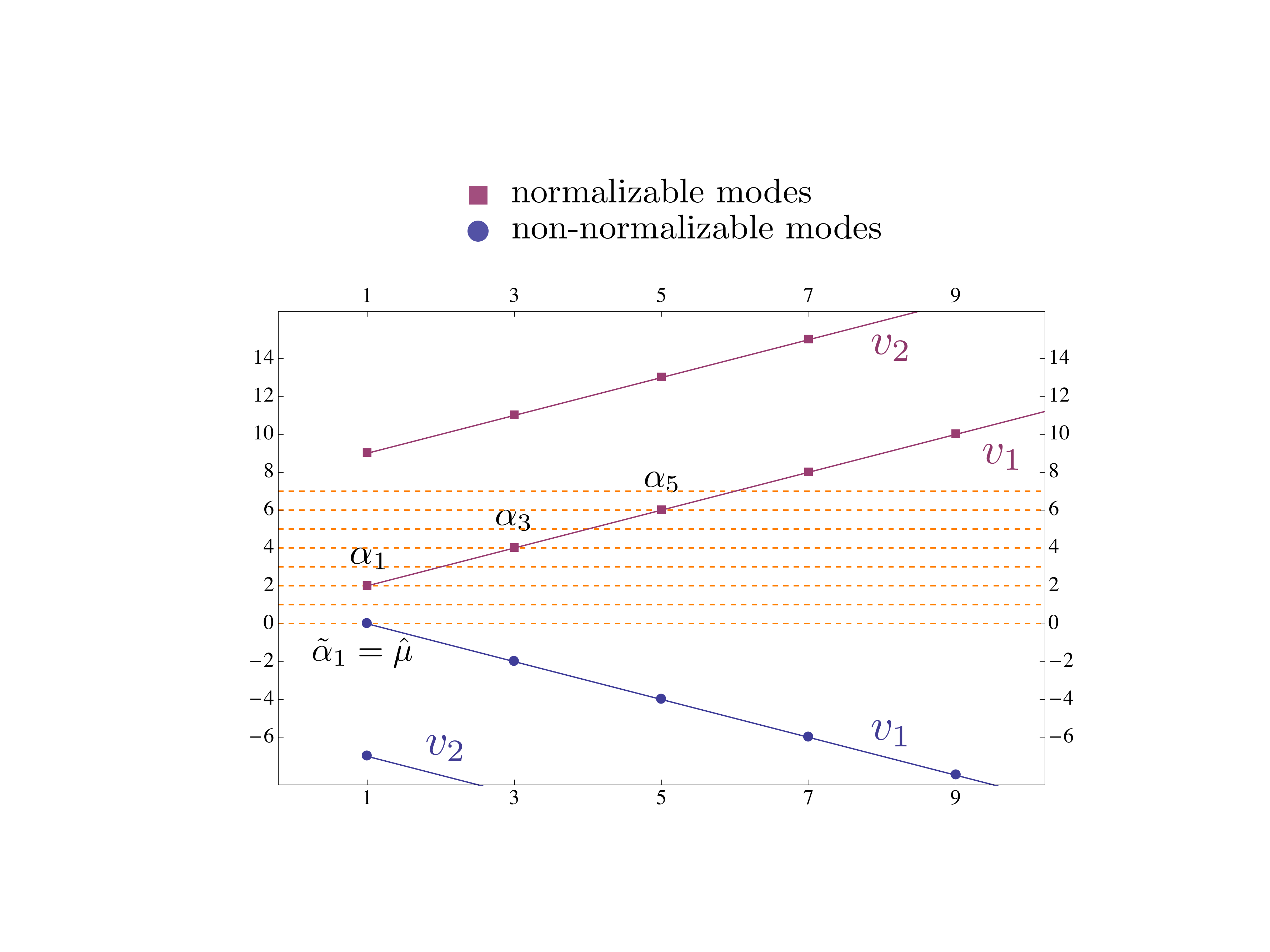}
\caption{\label{2-formPerturbations}
The  modes of the 2-form perturbation $v_1$ 
have a leading behaviour near the boundary of the form $y^{1-l}$ for non-normalizable modes
and  $y^{1+l}$ for normalizable. The figure shows the power of $y$  as a function of spin $l$. It also includes the powers for the other 
perturbation $v_2$. The dashed lines cover the region considered in this paper, up to order $y^7$.}
\end{figure}

Let us first consider the functions $M$ and $L$  that are associated to the  2-forms  on $S^8$  introduced above. The expansion in powers of $y$ can be seen to arise from the  normalizable and non-normalizable modes that are excited, plus their back-reaction. 
At the linear level there are two independent field perturbations associated to $M$ and $L$, which are called 
$v_1$ and $v_2$ in the perturbation analysis of \cite{Sekino:1999av}.  We can drop the perturbation $v_2$ because we impose that its  
non-normalizable modes  vanish,  and its normalizable modes  start at a power of $y$ beyond what we consider in this paper. 
Thus,  we have
\begin{align}
y^3 M(x,y) &=   \sum_{l\ge 1 \atop odd} \left( \alpha_l f^{(M)}_{l}(y) + \tilde{\alpha}_l \tilde{f}^{(M)}_{l}(y)\right)  \mathbb{H}_l(x) \,+ \, \text{back-reaction}\,,\nonumber \\
y^{-4} L(x,y) &=   \sum_{l\ge 1 \atop odd} \left( \alpha_l f^{(L)}_{l}(y) + \tilde{\alpha}_l \tilde{f}^{(L)}_{l}(y)\right)  \mathbb{H}_l(x) \,+ \, \text{back-reaction}\,,
\label{Q9andQ10expansion}
\end{align}
 where we denote non-normalizable modes with a tilde and normalizable without. These non-normalizable modes behave near the boundary as 
$\tilde{f}^{(M)}_{l}(y) \sim \tilde{f}^{(L)}_{l}(y) \sim y^{1-l}$. We set them all to zero but the mode $l=1$. This is the content of the boundary condition (\ref{Q9Q10_bc}), which sets
$\tilde{\alpha}_1=\hat{\mu}$, and defines the type of relevant deformation we decided to turn on.
Of course we are not free to set the normalizable modes to zero.  Their form can only be obtained once the solution is known everywhere, \emph{i.e.} once regularity deep in the bulk and at the axis is imposed.
These modes behave as  
$f^{(M)}_{l}(y) \sim f^{(L)}_{l}(y) \sim y^{1+l}$, near the boundary. Notice that the normalizable modes of the perturbations $v_2$,
which we dropped in (\ref{Q9andQ10expansion}), have
$f_{l}(y) \sim y^{8+l}$.
Figure \ref{2-formPerturbations} summarizes these facts.
In (\ref{Q9andQ10expansion}), we called back-reaction to all terms that are non-linear in the modes. At each order in the expansion at $y=0$, these can also be expanded in harmonic 2-forms on $S^8$. 
In this paper, we consider the first 8 terms in the expansion,
\begin{align}
&y^3M(x,y) = \hat{\mu} \left(1- \frac{9}{7} \,y^7\right)  \mathbb{H}_1(x) 
\label{expanQ9}
\\
&
-\frac{3}{176}
\hat{\mu }^3
   y^5 \big(43
   \mathbb{H}_1(x)-65
   \mathbb{H}_3(x)\big) 
    -\frac{3}{616}\alpha _1\hat{\mu
   }^2 y^7
   \big(97 \mathbb{H}_1(x)-350
   \mathbb{H}_3(x)\big) 
   -\frac{3}{4} \hat{\mu} (\delta +2\gamma)y^7
   \mathbb{H}_1(x) \nonumber\\
  &  -\frac{3}{50336}  \beta _2\hat{\mu
   }^3 y^7 \big(4811
   \mathbb{H}_1(x)-33488
   \mathbb{H}_3(x)+71148
   \mathbb{H}_5(x)\big) 
+O(y^8)\,.
\nonumber
\end{align}
\begin{align}
&y^{-4}L(x,y) = \hat{\mu} \left(\frac{3}{2}+ \frac{6}{35} \,y^7\right)  \mathbb{H}_1(x) 
+\alpha _1 y^2
   \mathbb{H}_1(x)+\alpha _3 y^4
   \mathbb{H}_3(x)+\alpha _5 y^6
   \mathbb{H}_5(x)
\label{expanQ10}
\\
& 
-\!\frac{315}{44}
   \beta _2 \hat{\mu } y^2
   \mathbb{H}_3(x)+\frac{1617
   }{80} \beta _4 \hat{\mu } y^4
   \mathbb{H}_5(x)
   +\frac{\hat{\mu }^3y^5}{880}
   \big(1575  \mathbb{H}_3(x)-464 \mathbb{H}_1(x)\big)-\frac{27027}{608} \beta _6 \hat{\mu
   } y^6 \mathbb{H}_7(x)\nonumber
\\
&  -\! \frac{3\hat{\mu }y^7}{80080}
 \big(
   26598 \alpha _1\hat{\mu }+47327 \beta
   _2 \hat{\mu}^2+52624 \gamma
   -4576 \delta
   \big)
   \mathbb{H}_1(x)
   \!+\!
   \frac{7\hat{\mu}^2 y^7}{1056}
  \big(528 \alpha_1+1165 \beta _2 \hat{\mu
   }\big) \mathbb{H}_3(x)\nonumber\\
   &-\frac{39053}{2496}  \beta
   _2 \hat{\mu }^3y^7
   \mathbb{H}_5(x)+O(y^8)\,.
\nonumber
\end{align}
where the coefficients $\beta_i$, $\delta$ and $\gamma$ appear in the expansion of scalar perturbations that we discuss
in a moment. 
Note that the first line in the expansions  (\ref{expanQ9}) and (\ref{expanQ10}) contains the terms linear in the modes, while the remaining terms arise from the back reaction of the fields.

\begin{figure}[t]
\centering
\includegraphics[height = 0.35\textheight]{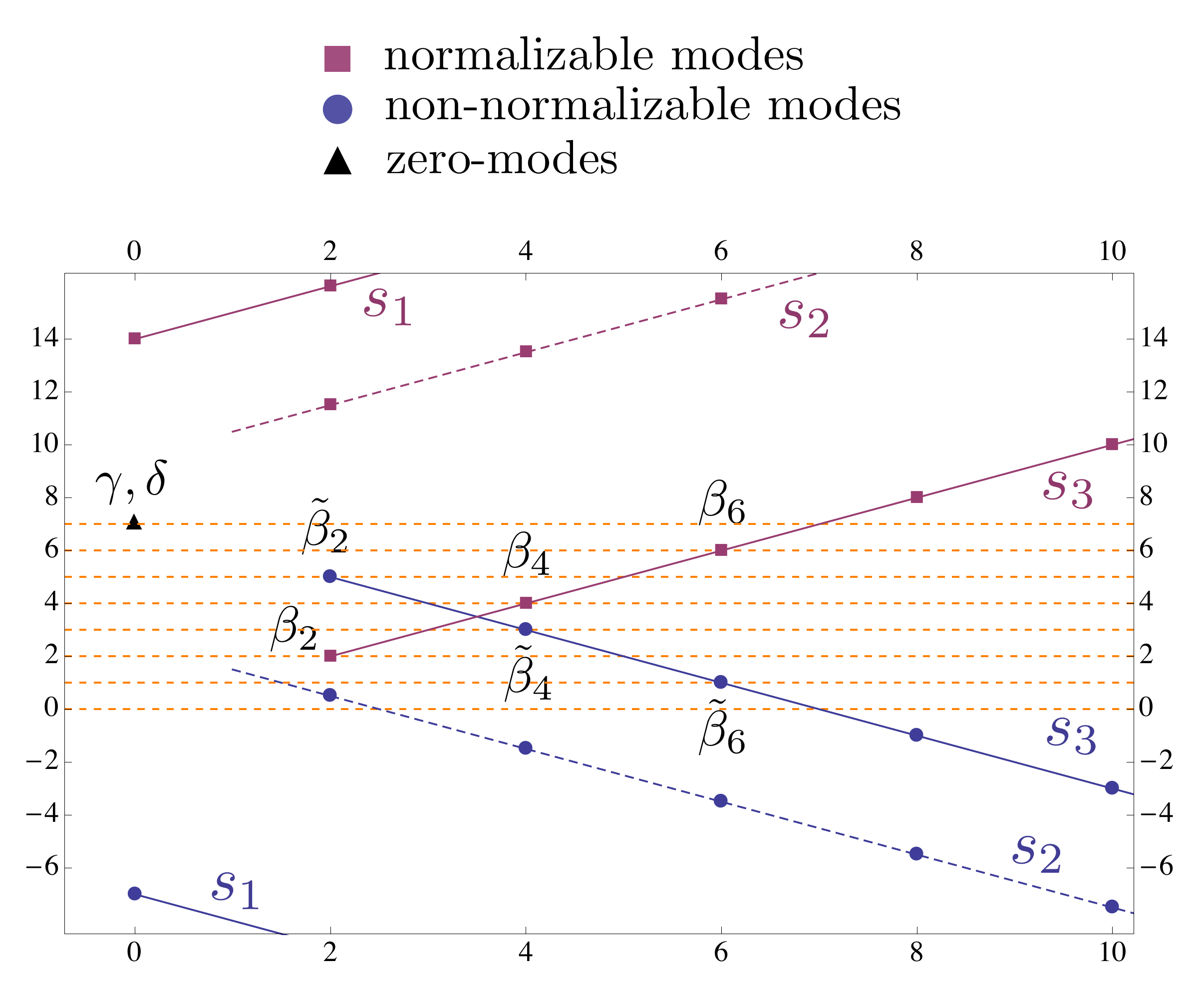}
\caption{\label{ScalarPerturbations}
Leading power of $y$ near the boundary of non-normalizable and normalizable modes of the scalar perturbation $s_3$.
Behaviour of the zero-modes is also included, as well as that of the scalar perturbations $s_1$ and $s_2$. The latter is  
dashed because it vanishes in the static case here considered.
Dashed horizontal lines cover the region considered in this paper.}
\end{figure}

The  modes of the 2-form perturbation $v_1$ 
have a leading behaviour near the boundary of the form $y^{1-l}$ for non-normalizable modes
and  $y^{1+l}$ for normalizable modes. Figure \ref{2-formPerturbations} shows the power of $y$  as a function of the spin $l$. It also includes the powers for the other 
perturbation $v_2$. The dashed horizontal lines cover the region considered in this paper, up to order $y^7$.

There are five  $S^8$ scalars in our Ansatz. They are  the functions $A$, $B$, $T_4$, $\Omega$ and the trace of the $S^8$ metric fluctuations
that we define by
\be
Q= \frac{T_2+2T_3+5T_4}{8}\,.
\ee
As in the previous case these fields can be expressed in terms of non-normalizable and normalizable modes, as well as the back reaction of all modes. 
In general, at the linear level there are three  independent scalar perturbations, called $s_1$, $s_2$ and $s_3$ in  \cite{Sekino:1999av}. The perturbation $s_2$
vanishes for our Ansatz (this follows because our geometry is static after reducing to type IIA supergravity). We also drop the perturbation $s_1$ because 
our boundary condition impose the vanishing of its non-normalizable modes and its normalizable modes only start at order $y^{14}$ in the expansion near the boundary.
Thus, for the purposes of this paper, we have 
\be
A(x,y) =   \sum_{l\ge 2 \atop even} \left( \beta_l f_{l}^{(A)}(y) + \tilde{\beta}_l \tilde{f}_{l}^{(A)}(y)\right)  \mathbb{S}_l(x) 
\,+ \, \text{zero-modes}\,+ \, \text{back-reaction}\,,
\label{Scalarexpansion}
\ee
and similarly for $B$, $T_4$, $\Omega$ and $Q$.
The functions $\mathbb{S}_l$ are $S^8$ scalar harmonics, that are $SO(3)\times SO(6)$ invariant, and  have the form
\be 
\mathbb{S}_l(x) = \frac{\mathrm{P}^2_{3+l}\left(x\sqrt{2-x^2}\right)}{15 x\sqrt{2-x^2}(1-x^2)^2} \,,
\ee
with $l$ even and $\mathrm{P}^m_n$ the associated Legendre polynomial of degree $n$ and azimuthal number $m$.
These functions are the usual eigenfunctions of the   Laplace operator on $S^8$ with 
\be
\star_8 d( \star_8 d \mathbb{S}_l ) = - \lambda_l \mathbb{S}_l\,, \quad\quad \lambda_l=l(l+7)\,.
\ee
We also impose that the non-normalizable modes of the perturbation $s_3$ vanish. As shown in figure \ref{ScalarPerturbations} 
this includes the three modes $\tilde{\beta}_2$, $\tilde{\beta}_4$, and
$\tilde{\beta}_6$ that appear in the expansion near the boundary at order $y^5$, $y^3$, and $y$, respectively. The leading behaviour of the normalisable modes
is also shown in the figure. The scalar perturbations also contain two zero-modes that we denote by $\gamma$ and $\delta$. These modes appear first
at order $y^7$  and are constant on the $S^8$. We call them zero-modes because they are not the zero frequency limit of any time dependent perturbation. 
Up to order $y^7$ the scalar perturbations have the form 
\begin{align}
&A(x,y)=1-\beta _2 y^2\mathbb{S}_2(x)
-\beta _4 y^4   \mathbb{S}_4(x)
   -\beta _6 y^6\mathbb{S}_6(x)- y^7\left(\delta +\frac{5}{2} \gamma\right)
   \mathbb{S}_0(x)  \\&
   +\frac{7}{143} \beta _2^2 y^4 \big(91   \mathbb{S}_0(x)+22
   \mathbb{S}_2(x)+30
   \mathbb{S}_4(x)\big)
    +\frac{1}{560} \hat{\mu }^2 y^5
   \big(50 \mathbb{S}_2(x)-21
   \mathbb{S}_0(x)\big)\nonumber\\&
 +\frac{14}{1105} \beta _2 \beta _4 y^6
   \big(612 \mathbb{S}_2(x)+220
   \mathbb{S}_4(x)+273
   \mathbb{S}_6(x)\big)
+  \frac{1}{840} \alpha _1 \hat{\mu }
   y^7 \big(495 \mathbb{S}_0(x)   +104\mathbb{S}_2(x)\big)\nonumber\\&
   +\frac{1}{205920}\beta _2 \hat{\mu }^2 y^7
   \big(-232245
   \mathbb{S}_0(x)+51582
   \mathbb{S}_2(x)+13600
   \mathbb{S}_4(x)\big)\nonumber\\&
   -\frac{49}{2431} \beta _2^3 y^6 \big(238
   \mathbb{S}_0(x)+561
   \mathbb{S}_2(x)+180
   \mathbb{S}_4(x)+126
   \mathbb{S}_6(x)\big)
    +O(y^8)\,,\nonumber
\end{align}
\begin{align}
&B(x,y)=1 
   +  \delta y^7 \mathbb{S}_0(x) 
  \\
   &+\frac{1}{2} \hat{\mu }^2 y^5
   \mathbb{S}_0(x)
     -\frac{16}{147} \alpha _1
   \hat{\mu }y^7
   \mathbb{S}_2(x)-\frac{5}{3003}  \beta _2
   \hat{\mu }^2 y^7\big(82
   \mathbb{S}_2(x)+35
   \mathbb{S}_4(x)\big)+O(y^8)\,,
   \nonumber
\\
&T_4(x,y)=1
 + \beta _2  y^2\mathbb{S}_2(x)
   + \beta _4  y^4 \mathbb{S}_4(x) 
   +\beta _6 y^6\mathbb{S}_6(x) 
    -\frac{7}{2}   \gamma y^7\mathbb{S}_0(x) 
\\    &   
   -\frac{1}{112} \hat{\mu }^2 y^5
   \big(7 \mathbb{S}_0(x)+18
   \mathbb{S}_2(x)\big)
    -\frac{1}{840} \alpha _1 \hat{\mu }
   y^7 \big(315 \mathbb{S}_0(x)+184
   \mathbb{S}_2(x)\big)\nonumber \\&
   -\frac{1}{41184} \beta _2 \hat{\mu }^2 y^7 \big(819
   \mathbb{S}_0(x)+24978
   \mathbb{S}_2(x)+22160
   \mathbb{S}_4(x)\big)
     +O(y^8)\,,\nonumber
\\
&\Omega(x,y)=1
-  \beta _2    y^2\mathbb{S}_2(x)
   -  \beta _4     y^4\mathbb{S}_4(x)
 - \beta _6 y^6 \mathbb{S}_6(x)
 +\frac{9}{14} \left( \gamma- \delta\right)   y^7\mathbb{S}_0(x)
\\&
 +\frac{7}{143} \beta _2^2 y^4 \big(91
   \mathbb{S}_0(x)+22
   \mathbb{S}_2(x)+30
   \mathbb{S}_4(x)\big) 
    +\frac{1}{560} \hat{\mu }^2 y^5
   \big(50 \mathbb{S}_2(x)-133
   \mathbb{S}_0(x)\big)\nonumber \\&
     -\frac{49}{2431}  \beta _2^3 y^6 \big(238
   \mathbb{S}_0(x)+561
   \mathbb{S}_2(x)+180
   \mathbb{S}_4(x)+126
   \mathbb{S}_6(x)\big)
   \nonumber \\&
   +\frac{14}{1105} \beta _2 \beta _4 y^6
   \big(612 \mathbb{S}_2(x)+220
   \mathbb{S}_4(x)+273
   \mathbb{S}_6(x)\big)
      +
 \frac{3}{280} \alpha _1 \hat{\mu }
   y^7 \big(35 \mathbb{S}_0(x)+16
   \mathbb{S}_2(x)\big)\nonumber \\&
   +\frac{1}{205920} \beta _2 \hat{\mu }^2 y^7
   \big(-257985
   \mathbb{S}_0(x)+39726
   \mathbb{S}_2(x)+13600
   \mathbb{S}_4(x)\big)
 +O(y^8)\,,\nonumber
\\
&Q(x,y)=1
+  \gamma  y^7  \mathbb{S}_0(x)
\\
&
-\frac{1}{16} \hat{\mu }^2 y^5
   \mathbb{S}_0(x)
   +\frac{2}{147} \alpha _1 \hat{\mu }
   y^7 \mathbb{S}_2(x)
   +\frac{1}{48048}  \beta
   _2 \hat{\mu }^2 y^7 \big(4054
   \mathbb{S}_2(x)+665
   \mathbb{S}_4(x)\big)
 +O(y^8)\,.\nonumber
\end{align}

Next let us consider tensor perturbations. These arise from the fields $T_1$, $T_2$ and $T_3$, which we can write as
\be
T_1\,\frac{4dx^2}{2-x^2}+T_2\,x^2(2-x^2) d\Omega_2^2+T_3
(1-x^2)^2 d\Omega_5^2\equiv
Q d\Omega_8^2+
T_{ij} d\theta^i d\theta^j\,,
\label{QandTij}
\ee
where $\theta^i$ denotes coordinates on $S^8$ and the symmetric tensor $T_{ij}$ is traceless with respect to the $S^8$ metric $h_{ij}$, i.e.
\be
h^{ij}T_{ij}=0\,,\quad \quad
d\Omega_8^2=
h_{ij} d\theta^i d\theta^j\,.
\ee
The trace part of these tensor perturbations is given by the function $Q$ already considered in the scalar perturbations above.
The modes that appear in the symmetric traceless tensor $T_{ij}$ can be divided in their divergence and divergence-less  parts.
The divergence part is obtained by acting on the scalar harmonics with the differential operator 
\be
\triangle_{ij} = \nabla_i\nabla_j - \frac{1}{8} h_{ij}  \Delta \,,
\ee
where $\Delta=\nabla^i\nabla_i$ is the $S^8$ Laplacian. These are, however, the same modes described
above for scalar perturbations. In fact, their appearance in the tensor perturbations can be gauge away by imposing the 
gauge condition $\nabla^i T_{ij}=0$. However, here they will be present in the tensor perturbations, since we do not
have such freedom, because in the DeTurck method a given gauge
choice is imposed on us.\footnote{Thus, a gauge transformation is necessary to make the precise map between our expansion and that of 
\cite{Sekino:1999av}.} 
The divergence-less part of these tensor perturbations comes again with non-normalizable and normalizable modes. We drop the 
non-normalizable modes and, for present purposes, we can neglect the normalizable modes because at linear level they appear first at order 
$y^9$. Working up to order $y^7$, the expansion of tensor perturbations reads  
\begin{align}
&T_{ij}(x,y)= 
  -\frac{1}{2352} \hat{\mu }^2 y^5 \big(16
   \Delta_{ij}
   \mathbb{S}_2(x)-945
   (\mathbb{T}_2)_{ij}(x)\big)+\frac{15}{28} \alpha _1 \hat{\mu }
   y^7 (\mathbb{T}_2)_{ij}(x)\\
   &-\frac{1}{144144} \beta _2 \hat{\mu }^2 y^7
   \big(968 \Delta _{ ij }
   \mathbb{S}_2(x)-110 \Delta
   _{ ij }
   \mathbb{S}_4(x)-24675
   (\mathbb{T}_2)_{ij}(x)+187110
  ( \mathbb{T}_4)_{ij}(x)\big)
 +O(y^8)\,,\nonumber
\end{align}   
where the $(\mathbb{T}_l)_{ij}$ are $S^8$ harmonic tensors that satisfy
\begin{align}
&h^{ij}(\mathbb{T}_l)_{ij}= 0\,,\quad\quad \nabla^i(\mathbb{T}_l)_{ij}=0\,,
\nonumber\\
&\Delta (\mathbb{T}_l)_{ij} = -\lambda_l(\mathbb{T}_l)_{ij}\,,
\quad\quad \lambda_l = l(l+7)-2\,,
\end{align}
with $l\ge 2$ and with $l$ even to guarantee invariance under the $SO(3)\times SO(6)$ subgroup.
Explicitly these harmonics are given by
\be
(\mathbb{T}_l)_{ij} d\theta^i d\theta^j =
\mathbb{R}_l(x)\,\frac{4dx^2}{2-x^2}+\mathbb{U}_l(x)\,x^2(2-x^2) d\Omega_2^2+\mathbb{V}_l(x)(1-x^2)^2 d\Omega_5^2\,,
\ee
where
\begin{align}
\mathbb{R}_l(x)=\ &  \,_2F_1\!\left(1-\frac{l}{2},\frac{9+l}{2},\frac{5}{2},x^2 \left(2-x^2\right)\right),
\\
\mathbb{U}_l(x)=\ & \left(1-8x^2 + 4x^4\right)\mathbb{R}_l(x) 
\nonumber\\
&-\frac{(l-2)(l+9)}{10} x^2 \left(2-x^2\right) \left(1-x^2\right)^2 \,_2F_1\!\left(2-\frac{l}{2},\frac{11+l}{2},\frac{7}{2},x^2 \left(2-x^2\right)\right),
\\
\mathbb{V}_l(x)=\ & -\frac{1}{5} \big( \mathbb{R}_l(x) + 2\mathbb{U}_l(x)  \big) \,.
\end{align}
Notice that, although an independent tensorial normalizable mode of spin $l$  appears first at order  $y^{7+l}$, these tensor perturbations 
already make their appearance at lower orders through the back-reaction.

Finally let us consider vector perturbations. For our Ansatz, $F$ is the  single $SO(9)$ vector. 
It turns out that there are no divergence-less vectors on $S^8$ that are  $SO(3)\times SO(6)$ invariant.
Thus the expansion of this field will only contain derivatives of the scalar perturbations, which is indeed 
confirmed by the expansion 
\begin{align}
y^{-1}F(x,y)&= 
\frac{1}{42} \hat{\mu }^2
     y^5\partial_x\mathbb{S}_2(x)
    +\frac{3}{98}
   \alpha _1 \hat{\mu }y^7  \partial_x\mathbb{S}_2(x) \\ &
   -\frac{1}{144144} \beta _2 \hat{\mu }^2 y^7
   \big(6862 \partial_x\mathbb{S}_2(x) -1155
   \partial_x\mathbb{S}_4(x) \big)
    +O(y^8)\,.\nonumber
\end{align}

All constants in the above expansions that remain to be determined correspond  to expectation values of dual operators in the matrix model. Up to order $y^7$ in the above expansions,
these are the constants $\alpha_1$, $\alpha_3$, $\alpha_5$ and $\beta_2$,  $\beta_4$,  $\beta_6$ and $\gamma$, $\delta$.   More normalizable modes show up at higher order, but we decided to only present results for these. 

For a more accurate numerical extraction of the remaining normalizable modes, we do a final change of variables that will ease the numerical procedure, namely we define 
\begin{align}
A &= 1+y^2 Q_1\,,\quad B = 1+y^5 Q_2\,,\quad 
F = 2 y^6 \sqrt{1-y} 
\left[\frac{\hat{\mu}^2}{84}\partial_x   \mathbb{S}_2(x)
+y\, Q_3 \right]\,,\\
 T_1 &= 1+y^5 Q_4\,,
 \quad T_2 = 1+y^5 Q_5\,,
 \qquad
 T_3 = 1+y^5 Q_6\,,
\quad T_4 = 1+y^2 Q_7\,,\quad \Omega = 1+y^2 Q_8\,,\nonumber
\\
M &= (1-y)y^{-3} \,\mathbb{H}_1(x)
\left[\hat{\mu}\,\frac{1-y^7}{1-y}+y^5 Q_9\right]\,,
\qquad
 L = \frac{3}{2}y^{4}\,\mathbb{H}_1(x)
\left[\hat{\mu}+y^2 Q_{10}\right]\,.
\nonumber
\end{align}
Our numerical procedure aims to solve for all ten $Q_i(x,y)$. We impose the following Neumann and Dirichlet  boundary conditions at $y=0$
\begin{align}
&\partial_y Q_1=\partial_y Q_{10}=Q_7+Q_1=Q_8-Q_1=0\ ,\qquad
Q_2=\frac{\hat{\mu}^2}{2}\ , \nonumber\\
&
Q_3=\frac{\hat{\mu}^2}{168}\partial_x   \mathbb{S}_2(x)\ ,\qquad
Q_9= \frac{3}{176} \hat{\mu}^3
 \left( 65 \frac{\mathbb{H}_3(x)}{\mathbb{H}_1(x) }
  - 43 \right)\,,
\label{finalBC}\\
&
Q_4\frac{4dx^2}{2-x^2}+Q_5x^2(2-x^2) d\Omega_2^2+Q_6
(1-x^2)^2 d\Omega_5^2 =
\frac{\hat{\mu}^2}{2351} 
 \big(945
   (\mathbb{T}_2)_{ij}(x) -16  \Delta_{ij}
   \mathbb{S}_2(x) \big) d\theta^i d\theta^j\,.
   \nonumber
\end{align}
These boundary conditions guarantee that all non-normalizable modes (except $\hat{\mu}$) are set to zero. In particular, the mode $\tilde{\beta}_2$ is the hardest to exclude because it only appears at order $y^5$ in the asymptotic expansion.
For example, a non-zero $\tilde{\beta}_2$  would give rise to
\be
Q_2(x,y)=\frac{\hat{\mu}^2}{2} +
\tilde{\beta}_2 \mathbb{S}_2(x) +O(y)\ .
\label{beta2tilde}
\ee
Therefore,  the boundary conditions (\ref{finalBC}) force $\tilde{\beta}_2=0$.

\subsubsection{Symmetry axes}

The boundary conditions at the equator $x=0$ are just those obtained via smoothness of the solutions. This implies that all $Q_i$ should be even functions of $x$, except $Q_3$, which should be odd under $x \to -x$. Moreover, we must have
$Q_4=Q_5$ at $x=0$ to avoid a conical deficit.
In practice, we just impose
\begin{equation}
Q_4(0,y)=Q_5(0,y)\ ,\qquad
Q_3(0,y)=0\  ,\qquad
\left.\partial_x Q_i(x,y)\right|_{x=0}=0\,,\quad \, i=1,2,5,\dots,10\,.
\end{equation}
Similarly, at the $x=1$ pole, we require that
$F$ is odd and  
\be
A,\ B,\ T_4,\ \Omega,\ M,\ L,\ \frac{T_1}{2-x^2},\ (2-x^2)T_2,\ (1+x)^2T_3
\ee
are even under  reflection around $x=1$.
Moreover, we avoid conical deficits by imposing
\begin{equation}
 Q_4(1,y)=Q_6(1,y)\,.
\end{equation} 
These conditions imply $\xi^x =0$ at $x=0$ and $x=1$.

At the horizon, which in the Euclidean setting is also a symmetry axis, regularity is easier to impose after changing to a new radial coordinate via $1-y=(1-\tilde{y})^2$. In the $\tilde{y}$ coordinate, the conditions for regularity are 
that $F$ is odd and all other functions are even under reflection around $\tilde{y}=1$. Moreover, we impose $A=B$ at  $\tilde{y}=1$
to avoid conical deficits with the periodicity $\Delta \eta = 4\pi/7$.
In practice, we use the boundary conditions
\begin{equation}
Q_1(x,\tilde{y}=1)=Q_2(x,\tilde{y}=1)\ ,\qquad
\left.\partial_{\tilde{y}} Q_i(x,\tilde{y})\right|_{\tilde{y}=1}=0\,,\quad \, i=2,3,\dots,10\,.
\end{equation}
These boundary  conditions   imply $\xi^{\tilde{y}}=0$.
 
It is a relatively easy exercise to show that the boundary conditions detailed above, together with the Einstein-DeTurck equations, form a well posed Elliptic problem \cite{2006math12647A,2007arXiv0704.3373A}. Furthermore, at the fictitious boundaries $x=0$, $x=1$, and $\tilde{y}=1$, the boundary conditions induced on $\xi$ are the relevant ones to admit $\xi=0$, \emph{i.e.} Einstein solutions, everywhere in the bulk \cite{Figueras:2011va}. We are thus ready to present our results and to detail the numerical method we used to solve the Einstein-DeTurck equations.

\subsection{Smarr formulae}

Having Smarr formulae is particularly important in situations in which the solution is presented numerically. Let us then recall how we can construct such formulae in the present case.
For every Killing vector $v$ of the solution (\ref{11Dansatz})
we can define an antisymmetric conserved tensor
\be
(K_v)^{ab}= \nabla^a v^b
+\frac{1}{4}G^{a bcd} v^e C_{cde}
+\frac{1}{18}v^{[a}G^{b]cde}  C_{cde}\ .
\ee 
Conservation of this tensor follows from the equations of motion (\ref{EinsteinEquation}), 
$\nabla^a G_{abcd}=0$, and from
 the identities
\be
\mathcal{L}_v g = \mathcal{L}_v C = \mathcal{L}_v G=0\ ,
\ \ \ \ \ \ \ \ \
\nabla_b\nabla^b v_a = -R_{ab} v^b\ .
\ee
In the language of differential forms, this means that we have a closed 9-form
\be
d \left(\star K_v\right) =0\,,
\ee 
where
\be
K_v = \frac{1}{2} (K_v)_{ab} \,dx^a \wedge dx^b\ .
\ee
Integrating $d (\star K_v)$ over a 10-dimensional surface $\Sigma_{12}$ of  constant time with $y_1<y<y_2$, we conclude that
\be
0= \int_{\Sigma_{12}}d \star K_v = 
\int_{\partial\Sigma_{12}}  \star K_v =
\int_{\Gamma(y_{2})}  \star K_v
-\int_{\Gamma(y_{1})}  \star K_v \ ,
\ee
where we used the fact that the boundary of $\Sigma_{12}$ has two disjoint components $\Gamma(y_{2})$ and $\Gamma(y_{1})$ (with opposite orientations).
This shows that the  integral
\be
I_v (y)=\int_{\Gamma(y)}  \star K_v \,,
\ee
over the closed surface of constant time and radial coordinate $y$ is independent of the value of $y$. The Smarr formula is obtained by equating the integral over the horizon $I_v (1)$ to the integral at infinity $I_v (0)$.

Choosing $v=\frac{\partial}{\partial \eta}$ to be the generator of time 
translations, we obtain
\be
\frac{7}{2} \hat{S} = 
\frac{8 \pi ^5}{3465}
 \left(924+2640 \gamma +594 \delta
  -220 \alpha _1 \hat{\mu }+119 \beta _2
   \hat{\mu }^2
 \right) .
 \label{SmarrT}
\ee 
To compute $I_v (1)$ we used the fact that the horizon is a Killing horizon of $\frac{\partial}{\partial \eta}$ with surface gravity equal to $\frac{7}{2}$. To compute $I_v (0)$   we used the  asymptotic expansion of the fields.

\begin{figure}[h]
\centering
\includegraphics[height = 0.3\textheight]{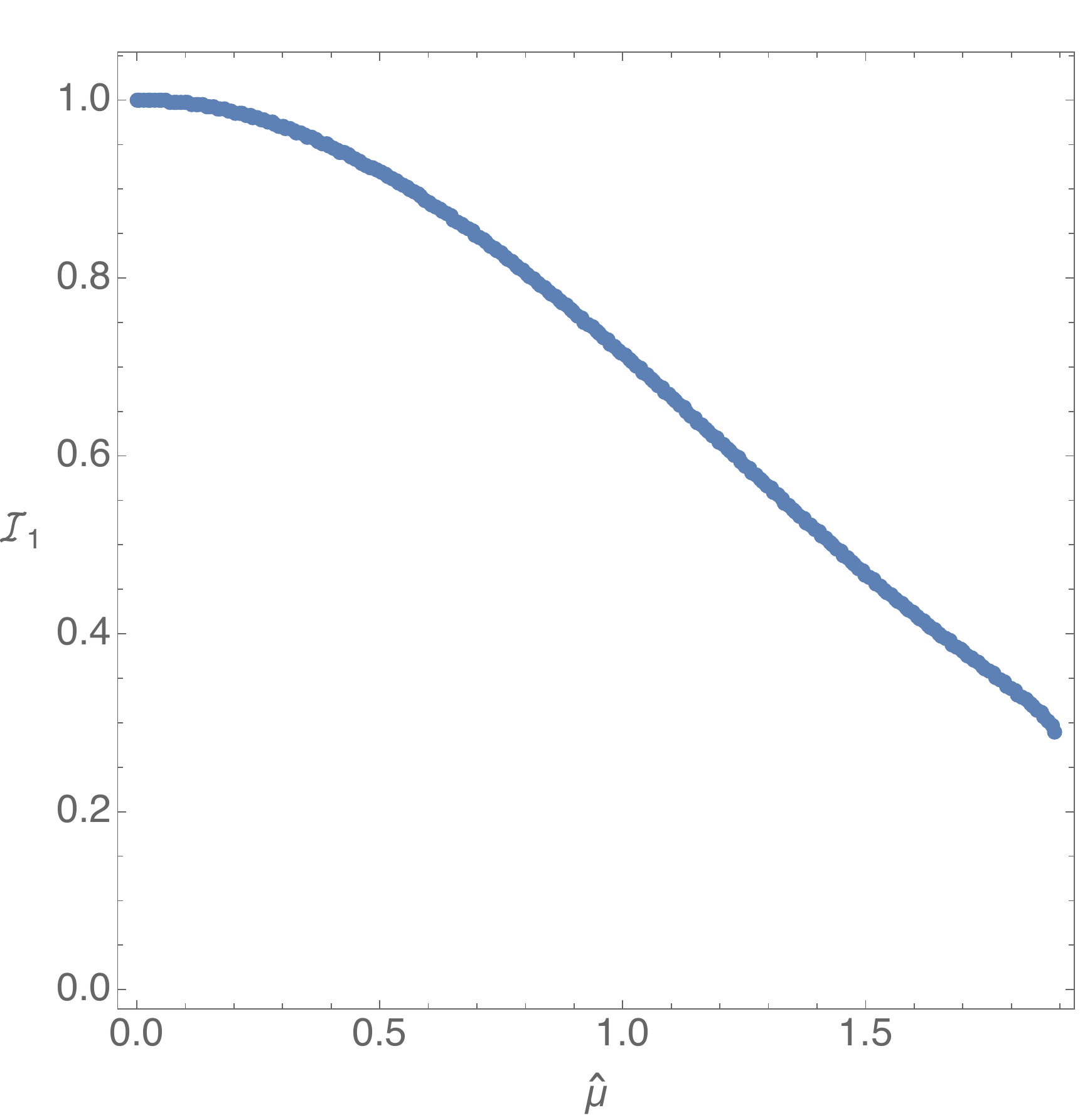}
\caption{\label{fig:BHcharge}
Momentum along the M-theory circle carried by the black string (first term in right hand side of (\ref{SmarrZ})). As $\hat{\mu}$ increases, the black string and the fields outside the horizon carry less and more momentum, respectively, keeping the total momentum of the geometry fixed.}
\end{figure}

Choosing $v=\frac{\partial}{\partial \zeta}$ to be the generator of  
translations along the M-theory circle, we obtain
\be
1=
\frac{105}{64\pi^5} \int_{\cal H} d^9x \sqrt{h}
\left(  \frac{T_4(1,x) \Omega(1,x)}{A(1,x)} 
-\frac{  L(1,x)\partial_yM(1,x)}{7x^4(2-x^2)^2A(1,x)T_2^2(1,x) } 
\right),
 \label{SmarrZ}
\ee
where the integration measure, given in (\ref{Shat}), is defined by the horizon metric.
This integral measures the momentum along the M-theory circle (or D0-brane charge in the type IIA picture) which is constant as we vary $\hat{\mu}$. We can think of the first term in (\ref{SmarrZ}) as the momentum carried by the black string, and the second term, which is also positive, is the momentum carried by the matter fields outside the horizon. 
In Fig. \ref{fig:BHcharge} we plot the momentum carried by the black string.
As $\hat{\mu}$ increases the momentum carried by the fields outside the horizon increases.  

It is also useful to integrate the $d(\star K_v)$
over the 10-dimensional   surface of constant $\zeta$. By a similar argument as the one above, we conclude that the following integral is 
independent of $y$
\be
\tilde{I}_v (y)=\int_{\tilde{\Gamma}(y)}  \star K_v \ ,
\ee
where $\tilde{\Gamma}(y)$ is the 9-dimensional surface of constant $y$ and $\zeta$.
Choosing $v= \frac{\partial}{\partial \zeta}$ we obtain $\tilde{I}_v (1)=0$ from the behaviour of the solution at the horizon. Thus, using  the behaviour  as $y\to 0$, we deduce the following identity 
\begin{equation}
\gamma +\frac{5}{44}\delta 
-\frac{1}{132} \alpha _1 \hat{\mu}
-\frac{287}{5808}\beta _2 \hat{\mu}^2=0\,,
  \label{identity}
\end{equation}
relating the parameters of the asymptotic expansion of the fields.

For $v= \frac{\partial}{\partial \eta}$ we also obtain $\tilde{I}_v (1)=0$. However,
 $\tilde{I}_v (0) $ depends on higher orders of $y$ in the asymptotic expansion of the fields than those considered above.

\subsection{Numerical solution}

We used a standard pseudospectral collocation in $x$ and $\tilde{y}$, and solved the resulting system of non-linear algebraic equations with a damped Newton-Raphson method. The dependence in $x$ and $\tilde{y}$ of all the functions was represented using tensor products of two Chebyshev collocation grids, each of which living on the unit interval $(0,1)$. Our integration domain is thus a square $(x,\tilde{y})\in(0,1)\times(0,1)$.

In expanding the functions $Q_i$ around the relevant boundaries, we have found no sign of non-smoothness. This means that a priori we expect the convergence of our method to be exponential in the number of grid points $\mathcal{N}$ and that no patching procedure is required. The only delicate numerical problem associated with these equations is that we need to accurately extract third and fourth derivatives off of the conformal boundary, in order to read the several constants corresponding to normalizable modes. For this reason, we decided to work with octuple precision and no less than $51$ grid points on each integration domain. In addition, due to the very bad condition numbers of the matrices we have to invert, we found useful to use up to twelve patches close to the boundary (depending on the values of $\widehat{\mu}$ and how steep our functions behave). These are conforming patches, which are patches that only coincide along a line, and have no overlapping regions. Since we are interested in accurately extracting asymptotic quantities, our patches coincide with lines of constant $\tilde{y}$ and cluster close to $\tilde{y} = 0$.

\begin{figure}[h]
\centering
	\begin{subfigure}{0.26\textheight}
		\includegraphics[height = 0.26\textheight]{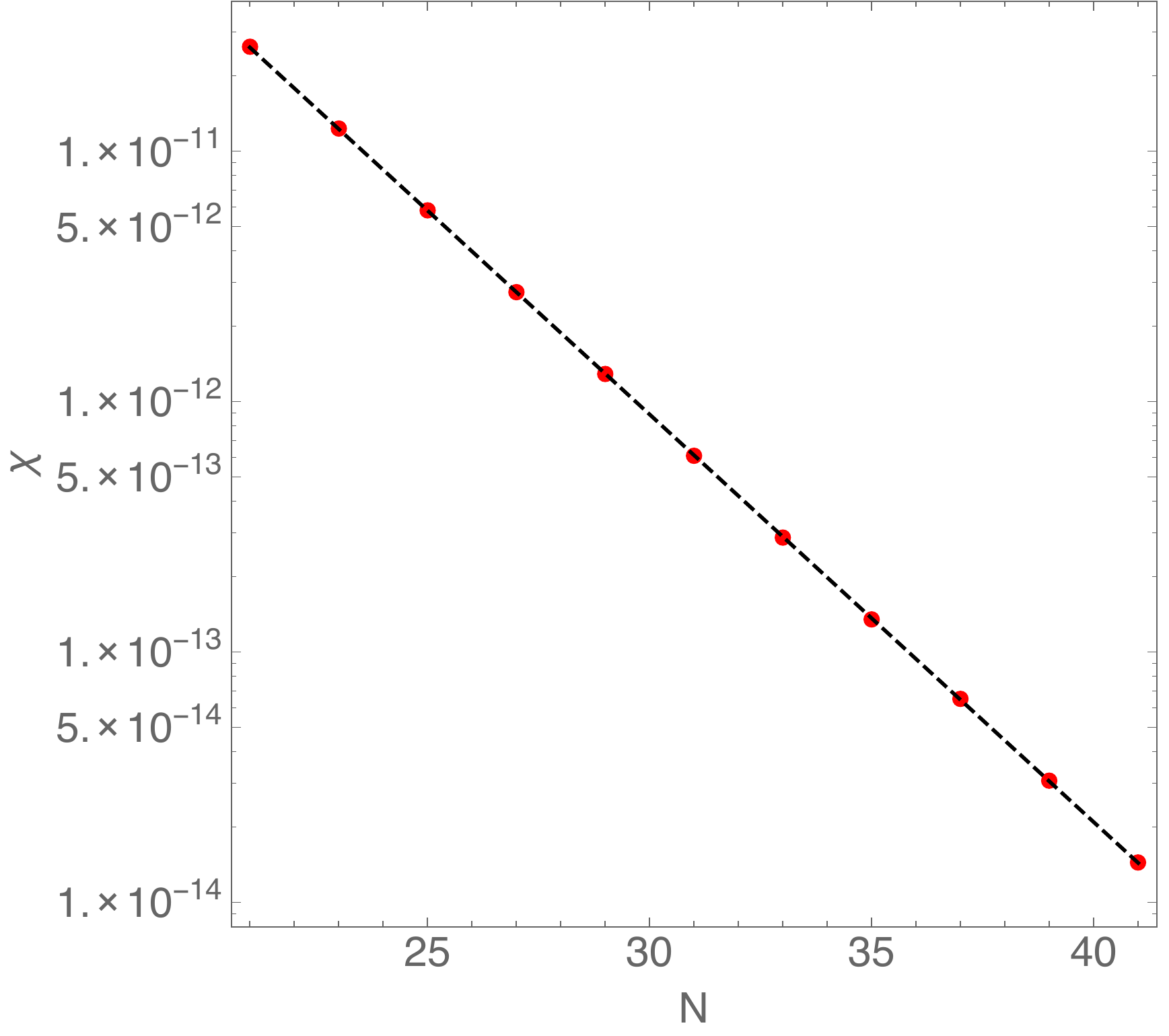}
		\caption{\label{fig:cona}Plot of $\chi$, as a function of $\mathcal{N}$.}
	\end{subfigure}
	\hspace{1 cm}
	\begin{subfigure}{0.26\textheight}
		\includegraphics[height = 0.26\textheight]{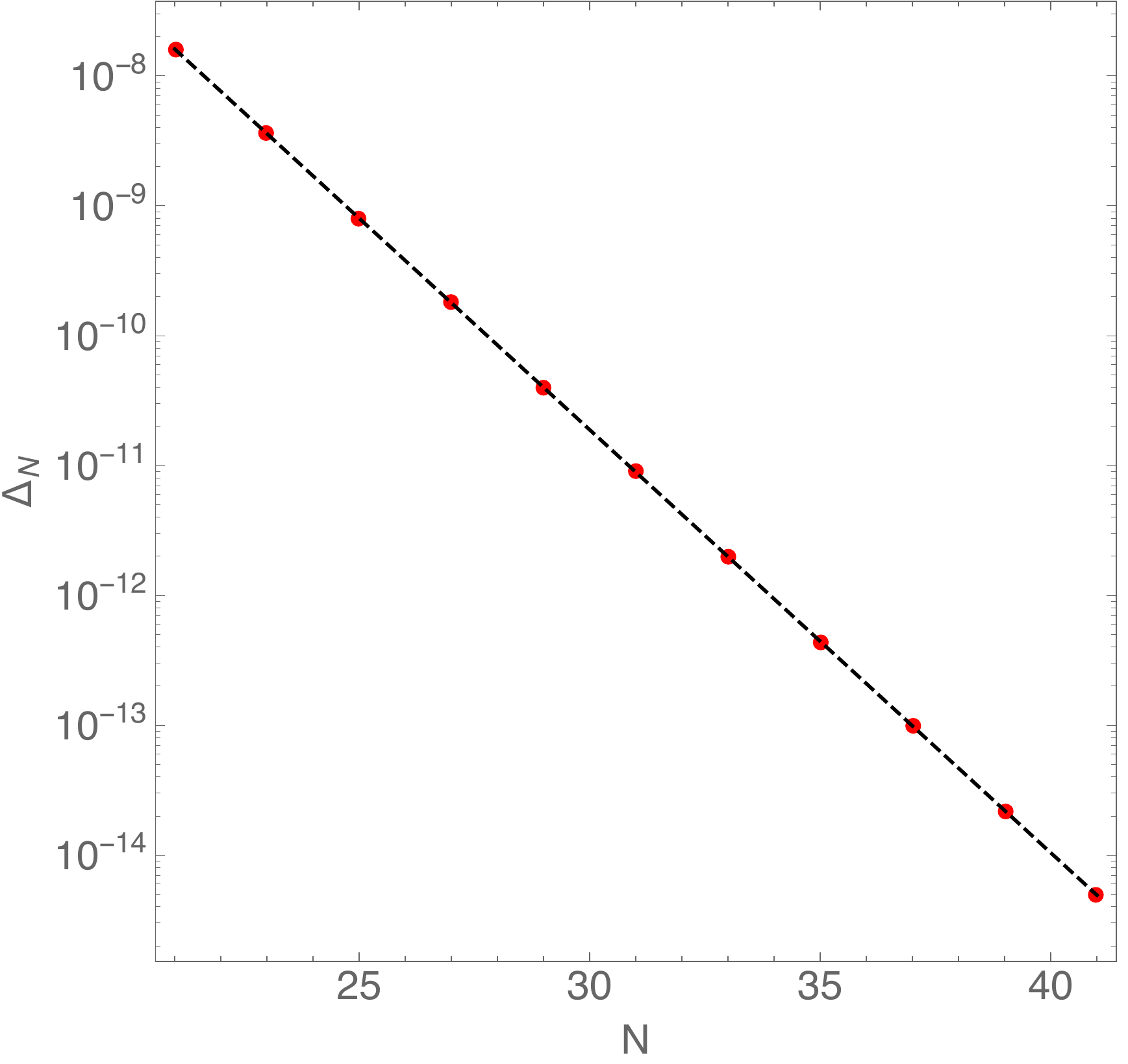}
		\caption{\label{fig:conb}Plot of $\Delta_{\mathcal{N}}$, as a function of $\mathcal{N}$.}
	\end{subfigure}
	\caption{\label{fig:con}Convergence plots for fixed $\widehat{\mu}= 1$.}
\end{figure}

In order to monitor the convergence of our numerical method, we monitored 
$\chi = \lVert\xi^a \rVert_{\infty}$ as a function of the number of grid points $\mathcal{N}$, as well as
\begin{equation}
\Delta_{\mathcal{N}} = \left|1-\frac{\widehat{S}_{\mathcal{N}}}{\widehat{S}_{\mathcal{N}+1}}\right|\,,
\end{equation}
where $\widehat{S}_{\mathcal{N}}$ denotes the entropy computed with $\mathcal{N}$ grid points in both directions. Both plots are displayed in Fig.~\ref{fig:con}, where a linear-logarithmic scale is used and we set $\widehat{\mu}=1$. The results are consistent with exponential convergence, as dictated by pseudospectral collocation methods.

A perhaps more striking test of our numerics comes from the identity (\ref{identity}).  We have checked that this relation is obeyed by our numerical data,  never exhibiting a violation above $10^{-6}\%$. 
Similarly, we checked that the Smarr formulas (\ref{SmarrT}) and (\ref{SmarrZ}) are verified by our numerical solutions with an accuracy of $10^{-6}\%$.  The Smarr formulae provide a very non-trivial validation  of  our numerical results because they relate quantities measured at the horizon ($y=1$) to quantities measured at infinity ($y=0$).
This  gives us full confidence that our numerical procedure is accurate enough for the physics we want to extract. 

\begin{figure}[h!]
\centering
\includegraphics[width = \textwidth]{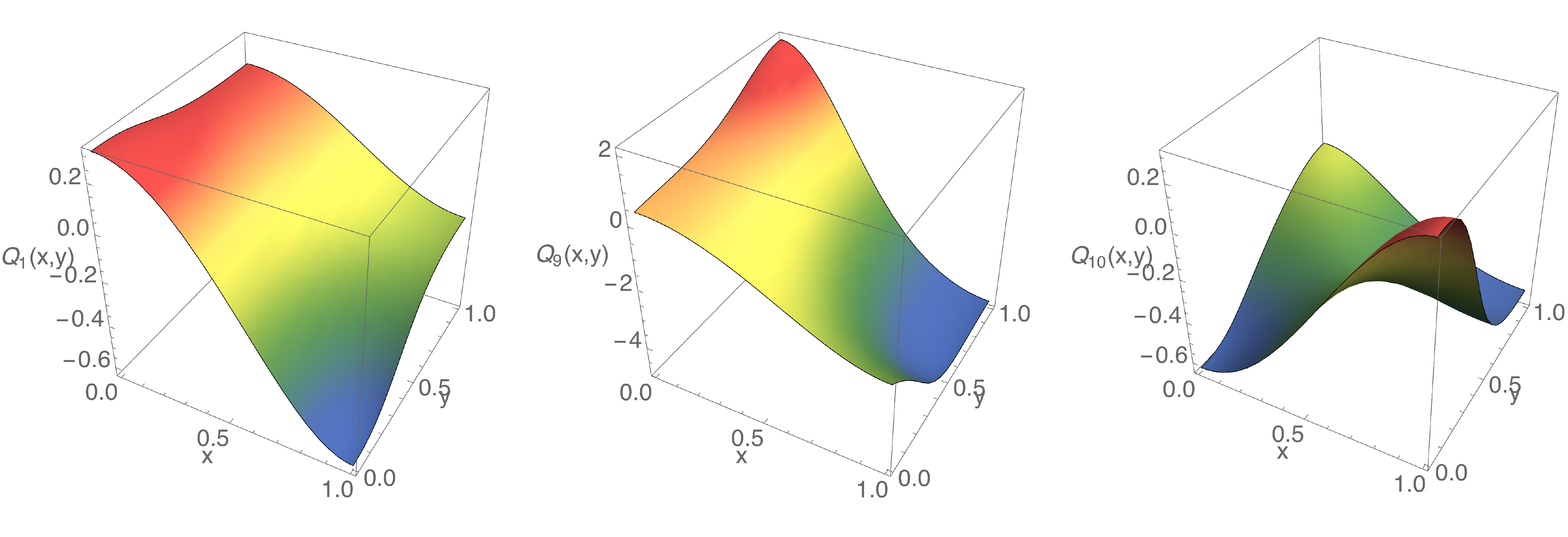}
\caption{\label{fig:example}From left to right: three-dimensional plots of $Q_1$, $Q_9$ and $Q_{10}$ as a function of $x$ and $y$ for fixed $\hat{\mu}=1$.}
\end{figure}

In Fig.~\ref{fig:example} we plot a typical run of our numerical method. It shows the behavior of $Q_1$, $Q_9$ and $Q_{10}$ as a function of $x$ and $y$. Note that these are all gauge invariant. From these plots we can easily see why we needed octuple precision, namely there is a large hierarchy between the functions. For instance, $Q_9$ evaluated on the horizon appears to be larger than all the remaining functions. This problem becomes worse as we increase $\widehat{\mu}$.

\begin{figure}[b!]
\centering
	\begin{subfigure}{0.26\textheight}
		\includegraphics[height = 0.26\textheight]{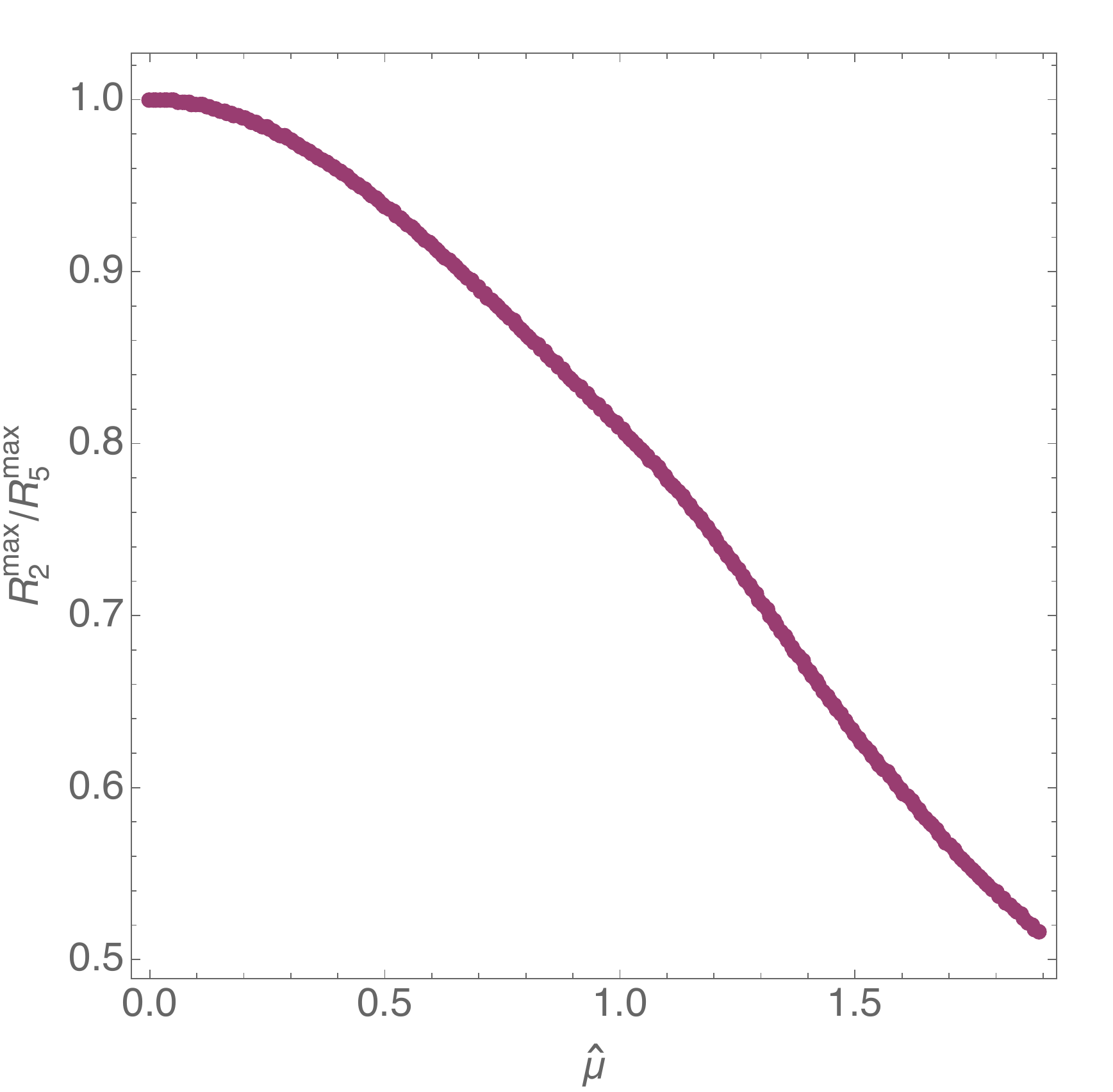}
		\caption{\label{fig:ratio}Ratio of the maximum size of the $S^2$, over the maximum size of the $S^5$, as a function of $\widehat{\mu}$.}
	\end{subfigure}
	\hspace{1 cm}
	\begin{subfigure}{0.26\textheight}
		\includegraphics[height = 0.26\textheight]{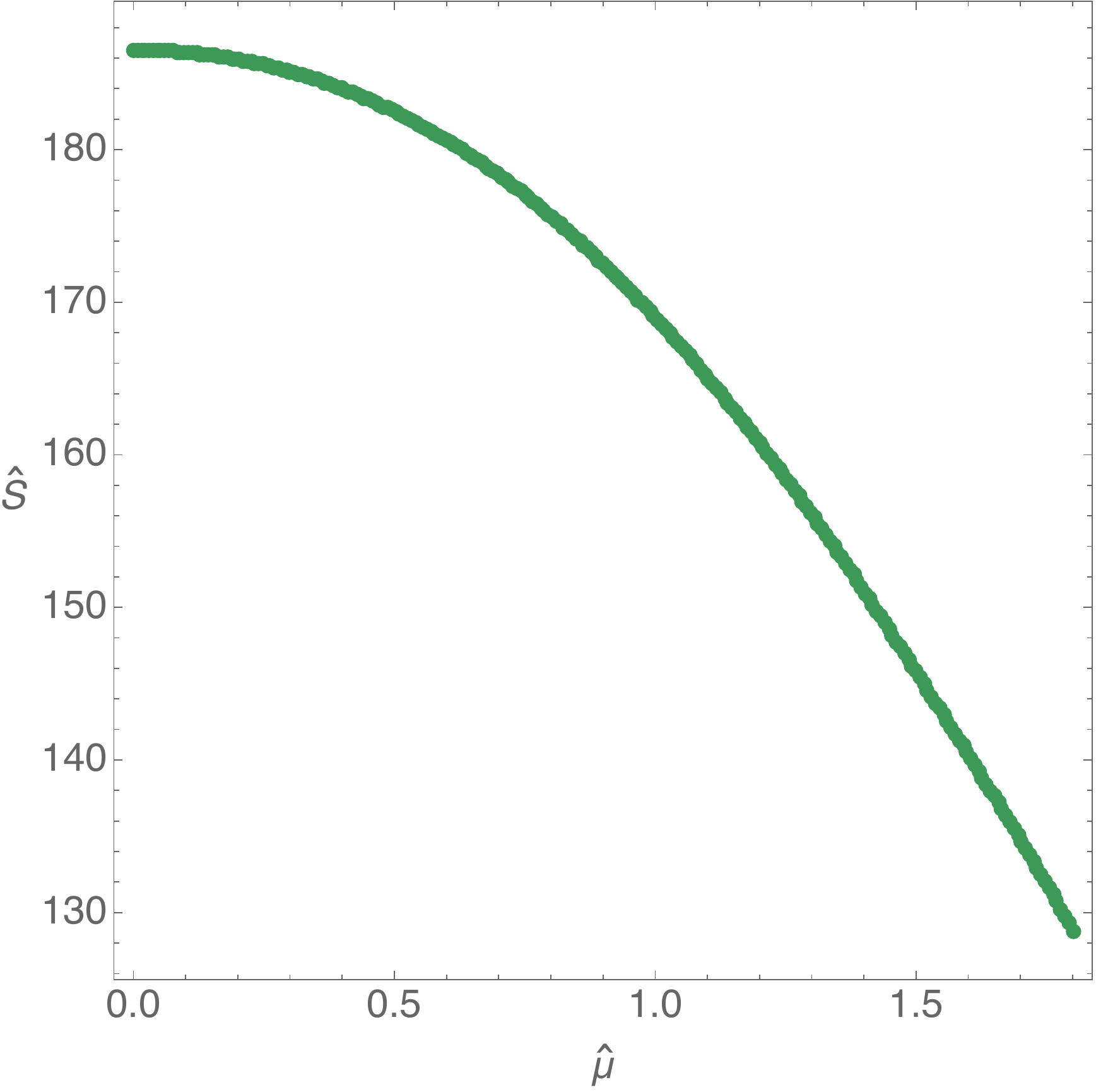}
		\caption{\label{fig:entropy}Normalized horizon area as a function of $\widehat{\mu}$.}
	\end{subfigure}
	\caption{Extracted physical quantities plotted as a function of $\widehat{\mu}$.}
\end{figure}

We now turn to more physical quantities. In particular, we would like to see how the horizon shape is changing as we change $\widehat{\mu}$. It is clear that the geometry will slowly move from having a round $S^8$ with $SO(9)$ symmetry to a deformed $S^8$  with a manifest $SO(3)\times SO(6)$. To explicitly quantify how deformed the horizon is from full spherical symmetry, we measure the radius of the $S^2$ at the pole and the radius of the $S^5$ at the equator. If the ratio between these quantities is very small, the horizon is highly distorted from spherical symmetry. We plot this quantity in Fig.~\ref{fig:ratio}.

The fact that this ratio reaches such small values might be worrying and suggestive of a Gregory-Laflamme type instability along the $S^5$ directions. In order to settle this, one would need to perturb this solution, and check its dynamical stability. We are currently undertaking this study, but have no results to report. Finally, we can also plot the normalized area of the horizon as a function of $\widehat{\mu}$, which we will need to reconstruct the free energy. This is done in Fig.~\ref{fig:entropy}, where we see the horizon area decreasing with increasing $\widehat{\mu}$. 

\begin{figure}[h!]
\centering
\includegraphics[height = 0.85\textheight]{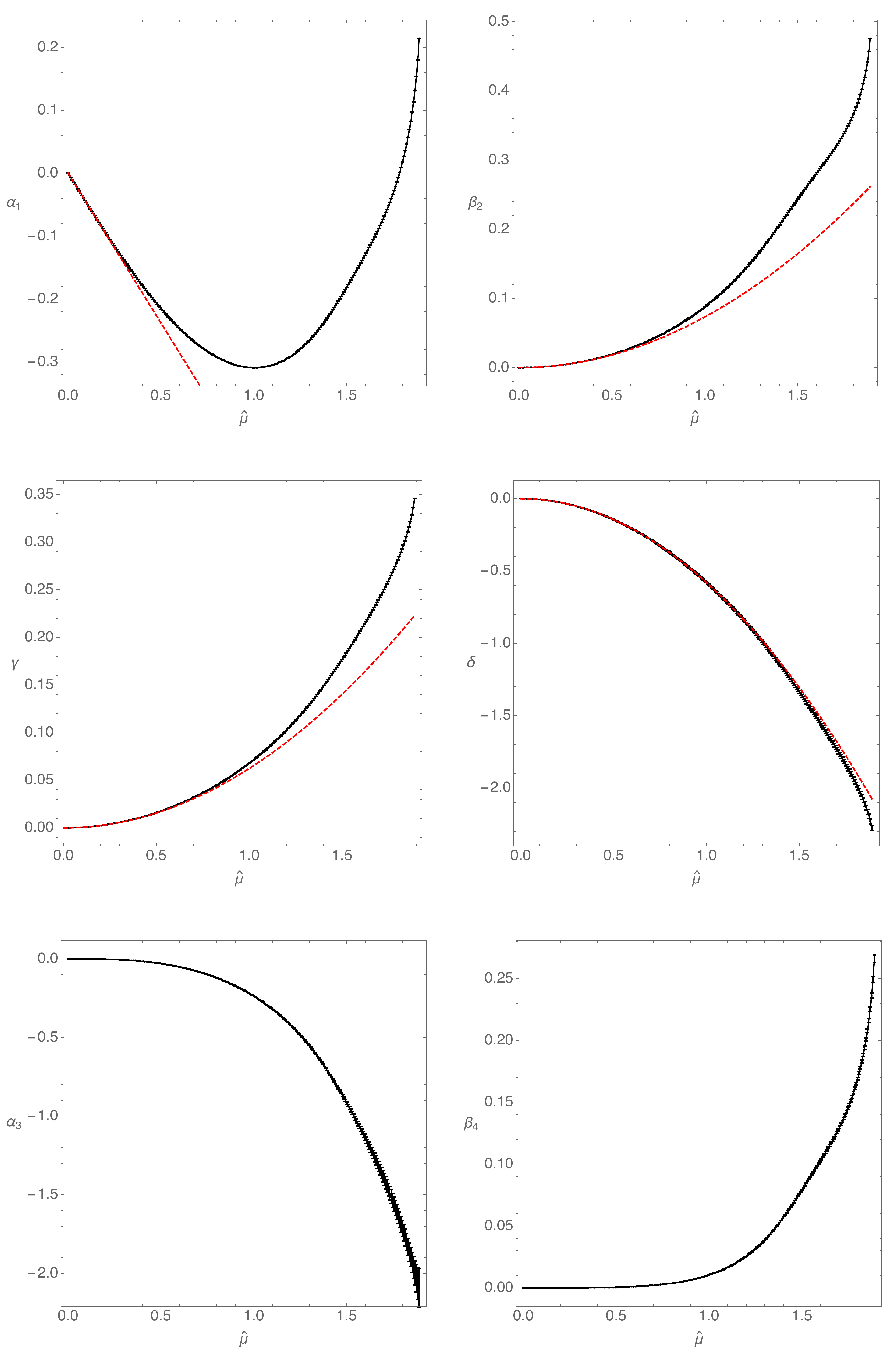}
\caption{\label{fig:panel}$\alpha_1$, $\beta_2$, $\gamma$, $\delta$, $\alpha_3$ and $\beta_4$ as a function of $\widehat{\mu}$. These plots contain error bars, obtained via a standard $\chi^2$ fit, which become increasingly larger as $\widehat{\mu}$ increases.
The red dashed lines are the prediction up to order $\hat{\mu}^2$, obtained from the study of perturbations around the $\hat{\mu}=0$ exact background ($\alpha_3$ and $\beta_4$ vanish to this order in $\hat{\mu}$). }
\end{figure}

We finalize this section by presenting, in Fig.~\ref{fig:panel}, the several extracted expectation values as a function of $\widehat{\mu}$. 
We obtain these expectation values by computing the first few $y$-derivatives at $y=0$ of the functions $Q_i(x,y)$ and fitting them to the asymptotic expansions discussed in section 
\ref{sec:asympAnsatz}.
A detailing of the predictions determined by perturbations around the $\hat{\mu}=0$ background (dashed red lines) is given in Appendix \ref{ap:pert}. Any Monte Carlo simulation of the PWMM (\ref{PWMMaction}) should hope to reproduce these results.

\section{Thermodynamics
\label{sec:thermo}}

Our numerical solution, expressed in terms of the functions in the Ansatz (\ref{11Dansatz}), depends on a single dimensionless
parameter $\hat{\mu}$ that determines the asymptotic behaviour of the 3-form potential $C$ through the boundary condition (\ref{Mbc}). Thus, 
the corresponding on-shell dimensionless action $\hat{I}$ and entropy $\hat{S}$, respectively defined in (\ref{physAction}) and (\ref{physEntropy}),
are functions of this single parameter. The boundary conditions imposed at the horizon fixed the periodicity of the Euclidean time circle
to $4\pi / 7$, independently of $\hat{\mu}$.  

Next, to obtain physical solutions from the above single-parameter family of solutions, 
we scaled the metric by $r_0^2$ and the 3-form $C$ by $r_0^3$, and changed the 
period of the M-theory circle according to (\ref{ScaleX11}). The new family of solutions,
parametrized by $\hat{\mu}$ and $r_0$, has the same leading asymptotics
of the non-extremal D0-brane solution (\ref{D011D}) with an additional  3-form potential $C$  with asymptotic
behaviour (\ref{asymptotics_dC}). It is then convenient to parametrize this new family of solutions
by the temperature $T$ and the mass deformation $\mu$, which are related to the original
single parameter by $\hat{\mu}=\frac{7}{12\pi}  \frac{\mu}{T}$  as derived in  (\ref{mu_hat}). Moreover, the 
on-shell action and entropy of the two-parameter and single-parameter families of solutions
are simply related by  (\ref{physAction}) and (\ref{physEntropy}), which we can rewrite in the form
\be
F(T,\mu) = -c_0 \,T^{\frac{14}{5}} \,\hat{I}\!\left(\hat{\mu}\right),\ \ \ \ \ \ \ 
S(T,\mu) =c_0\, \frac{14}{5} \, T^{\frac{9}{5}}\, \hat{S}\!\left(\hat{\mu}\right),
\label{Behaviour_F&S}
\ee
for a known (dimensionfull) constant $c_0$. In the particular case of zero mass deformation $\mu=0$ we 
recover the scaling with temperature as predicted directly from the matrix quantum mechanics in \cite{Wiseman2013, Morita:2013wfa}.
It is then clear that both the free energy and entropy are restricted to satisfy
\be
\frac{F(T,\mu)}{F(T,0)}
=\frac{\hat{I}\!\left(\hat{\mu}\right)}{\hat{I}(0)}\equiv f\!\left(\hat{\mu}\right),\ \ \ \ \ \ \  
\frac{S(T,\mu)}{S(T,0)}=\frac{\hat{S} \!\left(\hat{\mu}\right)}{\hat{S} (0)}\equiv s\!\left(\hat{\mu}\right),
\label{F&Srelations}
\ee
where, by definition, $f(0)=s(0)=1$.

\begin{figure}[h]
\centering
\includegraphics[height = 0.3\textheight]{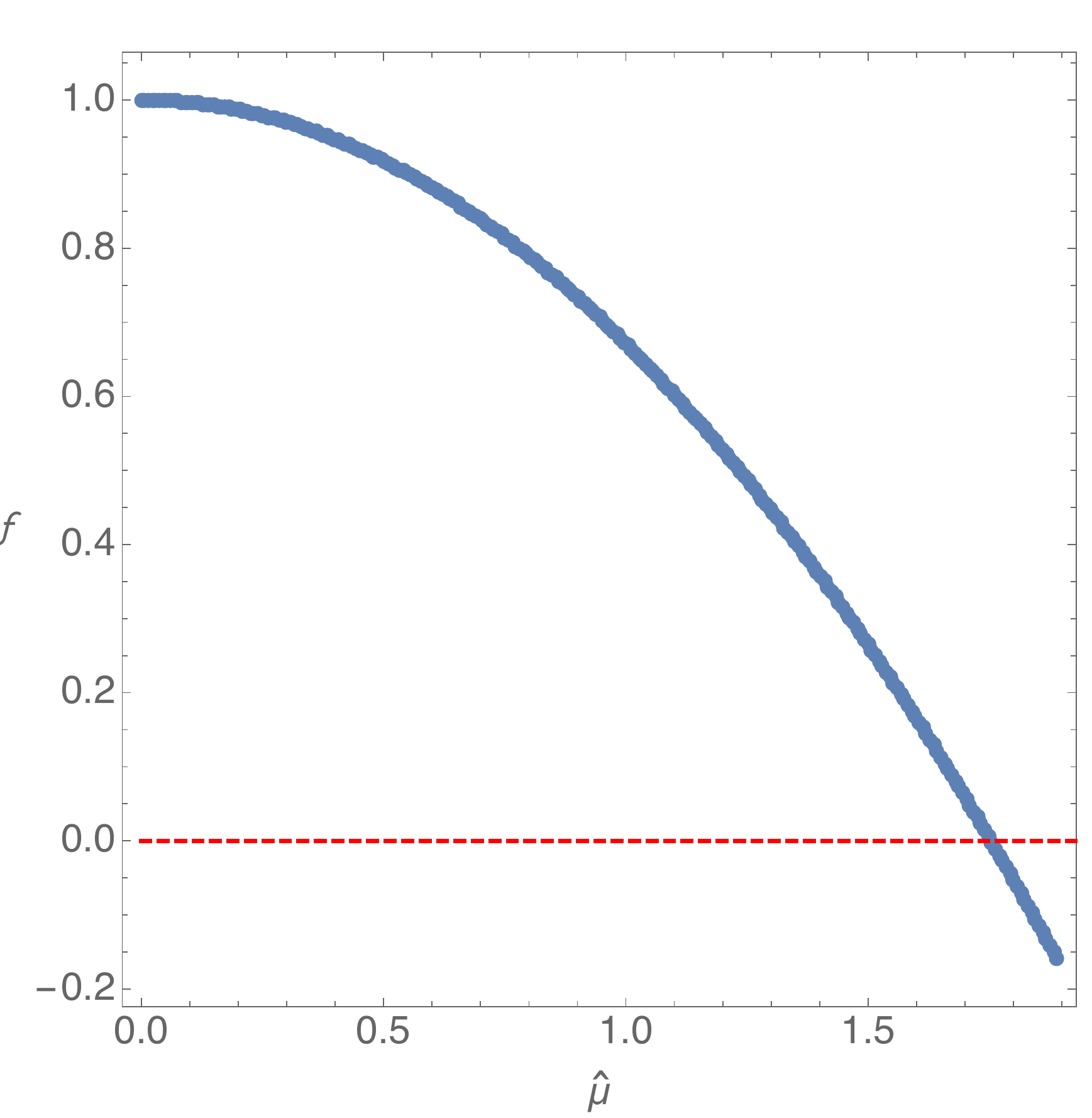}
\caption{\label{fig:free-energy}
The  free energy ratio $f(\hat{\mu})$ obtained numerically using (\ref{Seriesf}).}
\end{figure}

The behaviour of the free energy and entropy  (\ref{F&Srelations}), together with the 
scaling of the free energy as $T^{\frac{14}{5}}$ at zero mass deformation $\hat{\mu}$, can be used 
in the first law 
\be
\left(\frac{\partial F}{\partial T}\right)_\mu=-S\,,
\ee
to relate the functions
$f(\hat{\mu})$ and $s(\hat{\mu})$. This leads to the following equation
\be
\left(1-\frac{5}{14}\, \hat{\mu}\, \frac{\partial  }{\partial \hat{\mu} }\right) f(\hat{\mu})=s(\hat{\mu})\,,
\ee
which can easily be integrated
\be
f(\hat{\mu})=-\frac{14}{5}\,\hat{\mu}^{\frac{14}{5}}
\left[ C+ \int^{\hat{\mu}} dx\,x^{-\frac{19}{5}} s(x)\right],
\ee
where we wrote explicitly the  integration constant $C$.   
Notice that   the boundary condition $f(0)=1$  does not determine the constant $C$.
However, assuming that both $s(\hat{\mu})$ and $f(\hat{\mu})$ are analytic around $\hat{\mu}=0$, 
and therefore have a regular Taylor series expansion,
removes all ambiguity, 
\be
s(\hat{\mu})=\sum_{n=0}^\infty s_n \,\hat{\mu}^n\ \ \ \ \ \ \Rightarrow \ \ \ \ \ 
f(\hat{\mu})=\sum_{n=0}^\infty \frac{14 s_n}{14-5n} \,\hat{\mu}^n\,.
\label{Seriesf}
\ee
Since from computing the horizon area we know the function $s(\hat{\mu})$ numerically, 
we can do a polynomial fit to determine the first coefficients $s_n$, and then use it to plot $f(\hat{\mu})$   in Fig. \ref{fig:free-energy}.
The most important feature of this plot is that $f$ vanishes for $\hat{\mu}=\hat{\mu}_c\approx    1.7532672$. This means that, for $\hat{\mu}>\hat{\mu}_c$, the free energy of the deconfined phase of the PWMM is positive and of order $N^2$. Therefore, the confined phase that has a free energy of order $N^0$ will be smaller and dominate the thermal ensemble.
In other words, the critical temperature for the phase transition is
\footnote{We present the critical temperature with 6 digits  because our numerical solutions satisfied the Smarr formulas with $10^{-6}\%$ accuracy and the polynomial fit in (\ref{Seriesf}) decreases precision by one order of magnitude.}
\be
\frac{T_c}{\mu}= \frac{7}{12\pi \hat{\mu}_c}
= 0.105905(57)\,.
\ee

Let us now consider thermodynamical stability. 
 The specific heat of the system is given by
\be
c = 
T\left( \frac{\partial S}{\partial T}\right)_\mu.
\ee
From (\ref{Behaviour_F&S}) and (\ref{F&Srelations}) we may also express the specific heat in terms of the function
$s(\hat{\mu})$ as
\be
\frac{c}{S} = \frac{9}{5} -
\hat{\mu}\,\frac{\partial }{\partial\hat{\mu}}\log s(\hat{\mu})  \,.
\ee
Since in the range the black hole geometry is thermodynamically favoured,  $s(\hat{\mu})$ is a decreasing function, as shown in Fig. \ref{fig:entropy}, we conclude that the specific heat is always positive and therefore our solution is thermodynamically stable in this range.

\section{Discussion \label{sec:disc}}

Our main result is the construction of the black hole geometry dual to the deconfined phase of the PWMM. This allowed us to determine the value of the critical temperature at strong coupling as depicted in the phase diagram \ref{phasediagram}.
In addition, we determine the thermal expectation values of several observables in the deconfined phase (see Fig. \ref{fig:panel}).
 
At this point we would like to discuss an important caveat that we disregarded in the main text. There should be many black hole geometries with different horizon topologies and the same asymptotics as the solution we constructed. One can think of these as the finite temperature and backreacted versions of the many ways to distribute spherical probe M5 and M2 branes in equilibrium in the M-theory plane wave   \cite{BMN, Lin:2004kw}. These solutions are in one-to-one correspondence with the many vacua of the PWMM \cite{Maldacena:2002rb}. Our expectation is that the solution with lower free energy in the high temperature limit is the one we found because it has the simplest horizon topology. However, as we decrease the temperature it is possible that other black hole solutions start to dominate the thermal ensemble.\footnote{Such behavior has been observed at weak coupling in simulations of the PWMM \cite{Kawahara:2006hs}.}
Therefore, what we  really determined was  an upper bound for the critical temperature for the deconfinement transition. Notice that it is sufficient to find one black hole solution with negative free energy at a given temperature to conclude that the system must be in the deconfined phase at that temperature.
Even if this solution is dynamically unstable it must decay to another solution with lower free energy, thus the system remains in the deconfined
phase.

We hope our results motivate others to start a systematic exploration of the phase diagram of the PWMM by direct simulation of the matrix quantum mechanics, e.g. using the Monte-Carlo methods of \cite{ Catterall:2007fp,  Anagnostopoulos:2007fw, Catterall:2008yz, Hanada:2008ez, Hanada:2013rga}.
Our work provides concrete predictions for the behaviour of several thermodynamic quantities at strong coupling in the deconfined phase.
We also provide predictions for thermal expectation values of several operators.
However, the precise map between the gravitational parameters shown Fig. \ref{fig:panel} and operators of the PWMM is still missing.
This map is known \cite{Sekino:1999av} in the limit $\mu \to 0$ but its extension to finite $\mu$ requires the development of holographic renormalization with plane-wave asymptotics.

In fact, there has been a preliminary 
Monte-Carlo simulation of the PWMM \cite{Catterall:2010gf}. In this work, the authors simulate the PWMM at fixed temperature ($T/\mu=1/3$ in our conventions) and as they vary the coupling, they observe a first order phase transition for 
$0.03 \lesssim g \lesssim 0.045$.
This result is not in direct contradiction with our results but it implies a non-monotonic behaviour of the critical temperature as a function of the coupling $g$, complicating the phase diagram \ref{phasediagram}.
It would be great if this result could be confirmed by a more systematic Monte-Carlo simulation of the PWMM.

It would also be very interesting to perform a
Multicanonical Monte-Carlo simulation \cite{PhysRevLett.68.9, PhysRevLett.71.211, PhysRevE.64.056101} of the PWMM that could measure the density of states of the system.
This would provide a window into the thermodynamics of the system in the microcanonical ensemble, which is expected to have a richer structure, 
including a Hagedorn phase.\footnote{We thank Eliezer Rabinovici for emphasizing this point.}

Our black hole solution was constructed starting from the limit $\mu/T=0$. It would be interesting to understand our solution in the opposite limit $\mu/T \to \infty$. It is hard to address this question using our numerical methods because the black hole becomes very deformed and requires a much finer discretization grid.
In any case, Fig. \ref{fig:ratio} suggests that when $\mu \gg T$ the black hole looks like a pancake
(more precisely, a large 6D ball with a small thickness in the transverse 3 directions, times the M-theory circle). 
It should be possible to study this limit analytically using the blackfold approach of \cite{Emparan:2009at}.
The large deformation of the horizon also suggests that the system might be unstable to a  topology change to a ring-like horizon with $S^5\times S^3 \times S^1$ topology.
It should also be possible to study the low temperature regime of such black holes using the blackfold approach.
We leave these ideas for the future.


\section*{Acknowledgements}
We would like to thank Masanori Hanada, H\'elvio Vairinhos and Toby Wiseman for helpful discussions.  The research leading to these results has received funding from the [European Union] Seventh Framework Programme [FP7-People-2010-IRSES] and [FP7/2007-2013] under grant agreements No 269217, 317089 and No 247252, and from the grant CERN/FP/123599/2011
and from the Matsumae International Foundation in Japan. \emph{Centro de F\'isica do Porto} is partially funded by the Foundation for  Science and Technology of Portugal (FCT). 
M.S.C  and L.G. thank IPMU at Tokyo University for the great hospitality during the progress of this work.  L.G. is funded by the FCT/IDPASC fellowship SFRH/BD/51983/2012. This work was partially undertaken on the COSMOS Shared Memory system at DAMTP, University of Cambridge operated on behalf of the STFC DiRAC HPC Facility. This equipment is funded by BIS National E-infrastructure capital grant ST/J005673/1 and STFC grants ST/H008586/1, ST/K00333X/1.

\appendix
\section{Vacuum geometries \label{ap:LinMalda}}

The supergravity solutions dual to the vacua of the BMN model were constructed in \cite{Lin:2004nb, Lin:2005nh}.
They are given by
\begin{align}
ds^2=&\left(\frac{\dot{V}\Delta}{2V''}\right)^{\frac{1}{3}}\left[
-\frac{4 \ddot{V}}{\ddot{V}-2\dot{V}}dt^2+\frac{-2V''}{\dot{V}}(d\rho^2+dz^2)+
4d\Omega_5^2+2\frac{V''\dot{V}}{\Delta}d\Omega_2^2\right] \nonumber\\
&+\left(\frac{4 }{-V''\dot{V}^2\Delta^2 }\right)^{\frac{1}{3}}
\left(dx_{11} -\frac{2\dot{V}'\dot{V}}{\ddot{V}-2\dot{V}}dt\right)^2
\label{LMgeometries}\\
A_3=&-4\frac{\dot{V}^2V''}{\Delta}dt\wedge d^2\Omega_2+2\left(\frac{\dot{V}\dot{V}'}{\Delta}+z\right) dx_{11} \wedge d^2\Omega_2\ , \nonumber
\end{align}
where $\Delta=(\ddot{V}-2\dot{V})V''-(\dot{V}')^2$, the dot indicates derivative with respect to $\log \rho$ and the prime indicates derivative with respect to $z$.
The function $V(\rho,z)$ satisfies the Laplace equation in  cylindrical coordinates
\be
\frac{1}{\rho}\partial_\rho(\rho\partial_\rho V)+\partial_z^2 V=0\ ,
\ee
with asymptotic behaviour $V\approx \rho^2 z-\frac{2}{3} z^3 $ and boundary condition $V(z=0)=0$. A given vacuum of the BMN matrix quantum mechanics corresponds to a certain distribution of charged disks sourcing the potential $V$ as explained in \cite{Lin:2005nh}. 

For our purposes, it is sufficient to consider the asymptotic behaviour of the potential at large $z\sim \rho$,
\be
V=\rho^2 z-\frac{2}{3} z^3 + 
\frac{1}{60}
\sum_{l\,  {\rm odd}}  
\frac{2^{-l}\,a_l}{(\rho^2+z^2)^{\frac{l+1}{2}}}
P_l\left( \frac{z}{\sqrt{\rho^2+z^2} }\right)
\ee
where the $P_l$ is the Legendre polynomial and $a_l$ characterize each specific vacuum because they are  multipoles of the charge distribution that sources the potential $V$.
Inserting this expansion of the potential in the solution (\ref{LMgeometries}) we obtain an asymptotic expansion that can be compared with the asymptotic expansion of our Ansatz (\ref{11Dansatz}) discussed in section \ref{sec:asympAnsatz}.
More precisely, we perform the following change of coordinates in our Ansatz (\ref{11Dansatz})
\begin{align}
&d\eta= -\frac{dt}{  \sqrt{ a_1}}\ ,\qquad
d\zeta=2 \sqrt{ a_1} dx_{11}+ \phi_0 \frac{dt}{  \sqrt{ a_1}}\ ,\\
&y=\frac{1}{2\sqrt{\rho^2+z^2} }
\left[1+ \dots
\right], 
\qquad 
x=
\left(1-\frac{\rho}{\sqrt{\rho^2+z^2} }\right)^{\frac{1}{2}}
\left[1+ \dots
\right]
\nonumber
\end{align}
where the dots denote terms suppressed by powers of $\sqrt{\rho^2+z^2} $ that are determined so that our Ansatz (\ref{11Dansatz}) has the same type of asymptotic expansion as the vacuum solutions (\ref{LMgeometries}).
This comparison leads to the following relations
\begin{align}
&\hat{\mu}=2 \sqrt{ a_1} \ ,\qquad\qquad
\beta_2=\frac{a_3}{a_1}\ ,\qquad\qquad
\beta_4=\frac{a_5}{a_1}\ ,\qquad\qquad
\beta_6=\frac{a_7}{a_1}\ ,
\\
&\alpha_1=\frac{49 a_3}{33 \sqrt{ a_1}}\ ,\qquad\qquad
\alpha_3=-\frac{63 a_5}{10 \sqrt{ a_1}}\ ,\qquad\qquad
\alpha_5=\frac{2541 a_7}{152 \sqrt{ a_1}}\ ,\\
&
\gamma=\frac{5}{14}+\frac{149a_3}{396}+\phi_0\ ,\qquad\qquad
\delta=-\frac{22}{7}-\frac{136a_3}{99}-6\phi_0\ .
\end{align}
In other words, the vacuum geometries of \cite{Lin:2004nb, Lin:2005nh} have an asymptotic expansion of the form of section \ref{sec:asympAnsatz} with all the parameters given in terms the multipoles $a_l$ and an arbitrary constant $\phi_0$ that represents the freedom to shift the potential associated to the D0-brane charge (from the 10-dimensional point of view).
In addition, the parameter $\tilde{\beta}_2$, that appears for example in (\ref{beta2tilde}), vanishes in all vacuum solutions. This confirms our interpretation of $\beta_2$ as a state dependent response and $\tilde{\beta}_2$ as a source that deforms the theory.

\section{Perturbations Around the Background Solution \label{ap:pert}}

We begin by expanding the functions using  the spherical harmonics outlined in section 2,
\begin{align}
A &= 1+\hat{\mu}^2 y^2 \sum_l q_{1,l}(y)\mathbb{S}_l(x)+O(\hat{\mu}^4)\,,\qquad  \ B = 1+\hat{\mu}^2 y^2 \sum_l
q_{2,l}(y)\mathbb{S}_l(x)+O(\hat{\mu}^4)\,,\\
T_4 &=1+ \hat{\mu}^2 y^2 \sum_l q_{7,l}(y)\mathbb{S}_l(x)+O(\hat{\mu}^4)\,,\qquad \quad \Omega=1+\hat{\mu}^2y^2 
\sum_l q_{8,l}(y)\mathbb{S}_l(x)+O(\hat{\mu}^4)\,,\nonumber \\
F &= \hat{\mu}^2y^6\sqrt{1-y}\sum_l 
q_{3,l}(y)\partial_x \mathbb{S}_l(x)+O(\hat{\mu}^4)\,,\qquad
Q=1+\hat{\mu}^2y^5 \sum_l q_l(y)\mathbb{S}_l(x)+O(\hat{\mu}^4)\,, \nonumber \\
T_{ij} &=  \hat{\mu}^2y^5 \sum_l\left[ \tilde{q}_l(y)\Delta_{ij}\mathbb{S}_l(x)+\hat{q}_l(y)\mathbb{T}_{ij}(x)\right]+O(\hat{\mu}^4)\,, \nonumber \\
M &= \sqrt{1-y}\ y^{-3}\hat{\mu}
\sum_k q_{9,k}(y) \mathbb{H}_k(x)
+O(\hat{\mu}^3)\,,\qquad L = \frac{3}{2}y^{4} \hat{\mu}
\sum_k q_{10,k}(y)\mathbb{H}_k(x)+O(\hat{\mu}^3)\,,
\nonumber 
\end{align}
where $Q$ and $T_{ij}$ were introduced in (\ref{QandTij}).
We plug this Ansatz into the equations of motion and expand them up to $O(\hat{\mu}^2)$. 
In particular, the equation of motion $d\star dC=0$ gives rise to linear ODEs for the functions
$q_{9,l}(y)$ and  $q_{10,l}(y)$.
Moreover, the boundary conditions discussed in section \ref{sec:bc} imply that the only non-zero modes  are $q_{9,1}(y)$ and  $q_{10,1}(y)$.
We find these functions using a single variable version of the Chebyshev method described in the main text to reduce the two linear ODEs to a set of linear algebraic equations,   which can easily be solved using Newton's method (we  solve the equations in the $\tilde{y}$ coordinates because this simplifies the boundary conditions on the horizon). 


The harmonic Einstein equations $E_{ab}=0$, expanded up to $O(\hat{\mu}^2)$, can also be decomposed into spherical harmonics.
The equations of scalar type are $E_{\tau \tau}$,  $E_{yy}$,  $E_{zz}$, and  $E_{\tau z}$. For example
\be
E_{\tau \tau}(x,y)=\sum_l \mathbb{S}_l(x)
f_{1,l}(y).
\ee   
The vector equation $E_{yi}$ (where $i$ runs over the $S^8$ coordinates) can be decomposed as follows
\begin{align}
E_{yi}(x,y)=\sum_l \partial_i\mathbb{S}_l(x) f_{3,l}(y)\quad \Rightarrow \quad\nabla^iE_{yi}(x,y)=-\sum_l \lambda_l \mathbb{S}_l(x) f_{3,l}(y)\nonumber.
\end{align}
Finally, we can use the tracelessness of the tensor harmonics as well as the divergence-less nature of $\mathbb{T}_{ij}$ to write the components of the Harmonic Einstein equations corresponding to the $S^8$
\begin{align}
&E_{ij}= \sum_l \left(h_{ij}\mathbb{S}_l(x)f_l(y)+\Delta_{ij}\mathbb{S}_l(x)\tilde{f}_l(y)+\mathbb{T}_{ij}(x)\hat{f}_l(y)\right)\\
&\Rightarrow h^{ij}E_{ij}= 8\sum_l  \mathbb{S}_l(x)f_l(y)\,,\qquad
\nabla^i \nabla^j E_{ij}= -\sum_l \left[\lambda_l f_l(y)+\frac{7}{8}\lambda_l \left(1-\lambda_l \right)\tilde{f}_l(y)\right]\mathbb{S}_l (x)\,.\nonumber
\end{align}
Written in this way, the equations $E_{ab}=0$ can easily be projected onto a basis of the scalar harmonics, resulting in a system of seven ODEs, 
\be
f_{1,l}(y)=f_{2,l}(y)=f_{7,l}(y)=f_{8,l}(y)=f_{3,l}(y)=f_l(y)=\tilde{f}_l(y)=0\ .
\ee
These ODEs are linear in the functions $q_{1,l}(y),q_{2,l}(y),q_{7,l}(y),q_{8,l}(y),q_{3,l}(y),q_l(y),\tilde{q}_l(y)$ and quadratic in the functions $q_{9,l}(y)$ and $q_{10,l}(y)$ that can be previously determined from the gauge field equation of motion $d\star dC=0$.
\footnote{Note that the equation  $\hat{f}_l(y)=0$ is automatically satisfied setting $\hat{q}_l(y)=0$.}
The terms quadratic in the known functions $q_{9,l}(y)$ and $q_{10,l}(y)$ can be thought of as sources in the linear equations for the other 7 functions. 
This gives rise to 7 linear non-homogeneous ODEs   which can easily be solved using spectral methods.
To this order in $\hat{\mu}$, the sources are only non-zero for  $l=0$ and $l=2$.
%
%
%

The normalizable modes can be extracted from the solutions by comparing their behavior with the asymptotic expansions of the fields.
In particular,
\ba
\frac{\alpha_1}{\hat{\mu}}&=&\frac{3}{4} 
\partial^2_yq_{10,1}(0) +O(\hat{\mu}^2)\approx -0.4765+O(\hat{\mu}^2)\,,\\
\frac{\beta_2}{\hat{\mu}^2}&=&-q_{1,2}(0)+O(\hat{\mu}^2)\approx0.0732+O(\hat{\mu}^2)\,,\\
\frac{\gamma}{\hat{\mu}^2}&=&\frac{1}{2} \partial_{y}^2 q_{0}(0)+O(\hat{\mu}^2)\approx 0.0624 +O(\hat{\mu}^2)\,,\\
\frac{\delta}{\hat{\mu}^2}&=&\frac{1}{2} \partial_{y}^2 q_{2,0}(0)+O(\hat{\mu}^2)\approx-0.5809  +O(\hat{\mu}^2)\,.
\ea
%
%
These values satisfy the Komar identities (\ref{SmarrT}, \ref{identity}) and provide a non-trivial check of the numerics as shown in Figure 8. 

\bibliographystyle{utphys}
\bibliography{./mybib}
\end{document}